\documentclass[twocolumn,english,nofootinbib]{revtex4-1}
\usepackage[T1]{fontenc}
\usepackage[a4paper]{geometry}
\usepackage{babel}
\usepackage{graphicx}  
\usepackage{dcolumn}   
\usepackage{bm}        
\usepackage{amssymb,amsmath}   
\usepackage{verbatim}
\usepackage{amsmath,amssymb,amsfonts}
\usepackage{float} 
\usepackage{cleveref}              
\usepackage{blkarray}
\usepackage[usenames,dvipsnames]{xcolor}
\usepackage{epsfig}
\usepackage{epstopdf}
\usepackage{dcolumn}
\usepackage{tikz}
\usetikzlibrary{shapes.geometric, arrows}
\usepackage{upgreek}
\usepackage{setspace}
\usepackage{enumitem}
\usepackage{array,multirow,bigdelim}

\usepackage{tikz-cd}

\usepackage{moresize}

\usetikzlibrary{arrows,matrix,positioning}

\usetikzlibrary[decorations.markings]

\usepackage[compat=1.1.0]{tikz-feynman}
\usepackage{textgreek}

\def\mand{Mandelstam} 

\def\be{\begin{equation}}
\def\ee{\end{equation}}
\def\ba{\begin{eqnarray}}
\def\ea{\end{eqnarray}}

\def\a{\alpha}
\def\b{\beta}

\def\b#1{\overline{#1}}

\def\CP1{\mathbb{CP}^1}
\def\SL2C{\mathrm{SL}(2,\mathbb{C})}

\def\Z2{\mathbb{Z}_2}

\def\su2{{SU(2)}}
\def\eps{{\epsilon}}

\def\a{{\alpha}}

\def\[{\left[}
\def\]{\right]}

\def\L{\Lambda}

\def\s{\sigma}
\def\a{\alpha}
\def\b{\beta}

\def\[{\left[}
\def\]{\right]}

\def\<{\langle}
\def\>{\rangle}

\def\i2{\frac{i}{2}}

\def\2F1{\,_2{\rm F}_1}

\newcommand{\beq}{\begin{equation}}
\newcommand{\eeq}{\end{equation}}
\newcommand{\beqq}{\begin{equation*}}
\newcommand{\eeqq}{\end{equation*}}
\newcommand\beqa{\begin{eqnarray}}
\newcommand\eeqa{\end{eqnarray}}
\newcommand\beqaa{\begin{eqnarray*}}
\newcommand\eeqaa{\end{eqnarray*}}
\newcommand\bea{\begin{array}}
\newcommand\eea{\end{array}}

\newcommand{\ie}{{\it i.e. }}
\newcommand{\eg}{{\it e.g. }}
\newcommand{\etal}{{\it et al. }}

\begin{document}
\widetext
\title{A Novel On-Shell Recursive Relation}
\author{
Humberto Gomez
}
\affiliation{
Institute of Physics of the Czech Academy of Sciences \& CEICO \\ 
Na Slovance 2, 18221 Prague, Czech Republic. \smallskip \\
Facultad\! de\! Ciencias \! Basicas\! Universidad\!\\ Santiago\! de\! Cali,\!
Calle 5 $N^\circ$\!\! 62-00 Barrio Pampalinda\\ Cali, Valle, Colombia.
}

\begin{abstract}
We present a novel framework for deriving on-shell recursion relations, with a specific focus on biadjoint and pure Yang-Mills theories. Starting from the double-cover CHY factorization formulae, we identify a suitable set of independent kinematic variables that enables the reconstruction of amputated currents from amplitudes. As a byproduct, this new recursive structure recasts the BCJ numerators into an explicitly on-shell factorized form.
\end{abstract}

\maketitle

\section{Introduction}

The study of scattering amplitudes has transformed our understanding of quantum field theory, revealing elegant mathematical structures that remain hidden in traditional Feynman diagram techniques. This transformation centers on the idea that amplitudes (physical observables) can be directly constructed from on-shell, gauge-invariant quantities. The Britto-Cachazo-Feng-Witten (BCFW) recursion relations \cite{Britto:2005fq,Arkani-Hamed:2008bsc} are a seminal example of this paradigm, showing that tree-level amplitudes in gauge theories and gravity can be built from lower-point amplitudes. Through complex deformations of external momenta and the use of the global residue theorem, the BCFW relation bypasses the complications of off-shell intermediate states.

Another remarkable work in this direction is the Cachazo-He-Yuan formalism (CHY) \cite{Cachazo:2013gna,Mason:2013sva}, which recasts tree-level amplitudes as integrals over the moduli space of a punctured Riemann sphere. 
For example, the color-stripped $n$-point amplitude of the biadjoint massless scalar theory (BAS) in the single-cover formulation (SC) is given by the integral,
\begin{equation}\label{BAchy}
A^{\phi^3}_n(\a|\beta) = \int_\gamma \, \prod_{a=3 \atop a\neq p,q,r}^{n-1} \frac{d\sigma_a}{S_a} \,\Delta_{(pqr)}^2\,
{\rm PT}(\a)\, {\rm PT}(\beta) ,
\end{equation}
where $\alpha$ and $\beta$ are two specific orderings, $\sigma_a$ is the position of the $a^{\rm th}$ puncture on the Riemann sphere, and $\gamma$ is the integration contour defined by the equations, $S_a=0$, for $a=3,\ldots, n-1$. The $S_a$'s, known as the ``scattering equations'', are given by, 
\begin{equation}
S_a \equiv \sum_{b=1, b\ne a}^n \frac{2\, k_a \cdot k_b}{\sigma_{a}-\sigma_{b}}=0 ,
\end{equation}
where $k_a$ is the momentum associated with the $a^{\text{th}}$ external (ordered) particle. We have also introduced the notation, $\Delta_{(pqr)}\equiv \s_{pq}\s_{qr}\s_{rp}$, with $\s_{ij}\equiv \s_i -\s_j$, and ${\rm PT}(\a)\equiv(\s_{\a_1\a_2}\s_{\a_2\a_3}\cdots \s_{\a_{n}\a_1})^{-1}$. 
Without loss of generality, the ${\rm SL}(2,\mathbb{C})$ symmetry has fixed by setting the punctures $(\sigma_p, \sigma_q, \sigma_r)$ to constants. 
Here, the momentum conservation, \(\sum_{a=1}^n k_a = 0\), and massless condition, $k_b^2=0$, are implicit.

When at least one of the external legs is off-shell, \ie $k_b^2 \neq 0$, we refer to the resulting object as an \textit{amputated current}, denoted by the calligraphic symbol ${\cal J}^{\phi^3}_n$ \cite{Naculich:2015zha,Gomez:2016bmv,Bjerrum-Bohr:2018lpz,Cachazo:2021wsz}.
In the special case when $\alpha = \beta$, we adopt the shorthand notation,
\begin{equation}
A^{\phi^3}_n(\a) \equiv	 A^{\phi^3}_n(\a|\a), \quad 
{\cal J}^{\phi^3}_n(\a) \equiv	 {\cal J}^{\phi^3}_n(\a|\a).
\end{equation}
The amplitudes $A^{\phi^3}_n(\alpha)$ correspond to the theory known as ${\rm Tr}(\phi^3)$ \cite{Arkani-Hamed:2024fyd}.
 
A refined version, the double-cover formalism (DC) \cite{Gomez:2016bmv}, offers a new handle that enables a clearer understanding of factorization properties  \cite{Gomez:2016bmv,Bjerrum-Bohr:2018lpz}. In this formulation, the ${\rm SL}(2,\mathbb{C})$ symmetry is extended to ${\rm GL}(2,\mathbb{C})$, allowing us to gauge-fix four punctures instead of three. Consequently, one of the ``scattering equations'' becomes non-zero and instead manifests as a propagator.

In this work, we begin by studying the factorization formulae obtained from the double-cover approach \cite{Gomez:2016bmv,Gomez:2016bmv,Bjerrum-Bohr:2018lpz}, with a focus on biadjoint and pure Yang-Mills (YM). Follows, we identify a suitable set of independent kinematic variables for which the amputated currents (emerging from the double-cover CHY factorization formulae) are manifestly independent of the Mandelstam variables associated with the ``mass'' of the off-shell legs. This invariance allows us to set those variables to zero, thereby reducing the computation to purely on-shell data, \ie amplitudes. 
Finally, given a profound understanding of ordered theories is encapsulated in the Bern-Carrasco-Johansson (BCJ) numerators \cite{Bern:2008qj}, we show that our new on-shell formulation naturally yields a factorized structure for the BCJ numerators.

It is important to emphasize that in this work, the number of spacetime dimensions is not specified. Instead, we regard the kinematic space defined solely by the Mandelstam invariants $s_{ij}=(k_i+k_j)^2$. These variables are not all independent, as momentum conservation imposes the constraints $\sum_j s_{ij} = 0$. Consequently, a scattering process involving $n$ massless particles has $\frac{n(n-1)}{2} - n = \frac{n(n-3)}{2}$ independent kinematic invariants.

\section{Recursion for BAS from the DC formulation}\label{}

In the double-cover (DC) formulation of the CHY construction, the $n$-point amplitude is expressed as a contour integral over a double-covered Riemann sphere with $n$-punctures.
Schematically, the pairs $(y_1,\sigma_1), (y_2, \sigma_2),\ldots, (y_n,\sigma_n)$ define a new set of integration variables, constrained to lie on the algebraic curve,
\noindent
\begin{align}
	C_a \equiv y_a^2 - \sigma_a^2 + \Lambda^2=0 \qquad \textrm{for } a = 1,\ldots,n.
\end{align}
which encodes the double-cover geometry.
A translation table has been worked out in detail in \cite{Gomez:2016bmv}. 

The DC approach for the color stripped BAS amplitude for two different orderings, $\a$ and $\b$, is given by the integral (let us recall that \(\sum_{a=1}^n k_a = 0\), and $k_b^2=0$),
%
\begin{equation}\label{BAdouble}
A^{\phi^3}_n(\a|\beta)  = \int_\Gamma d\mu_n \, 
 \frac{\Delta^\tau_{(pqr)} \, \Delta^\tau_{(pqr|m)}} {S^\tau_m } \,
{\rm PT}^\tau(\a)\, {\rm PT}^\tau(\b) ,
\end{equation}
where, without loss of generality, the ${\rm GL}(2,\mathbb{C})={\rm SL}(2,\mathbb{C})\times \mathbb{C}^*$ symmetry has fixed by setting the punctures $(\s_p,\s_q,\s_r,\s_m)$ to constants. 

 The scattering equations, measure, and Parke-Taylor factor are given by the expressions
\begin{align}
& \tau_{(a,b)}  \equiv \frac{1}{2\,\s_{ab}}\left(\frac{y_a+y_b+\sigma_{ab}}{y_a}\right), \,\,  \nonumber \\
& S^\tau_a  \equiv 
\sum_{b=1 \atop b\neq a}^n  2 \, k_{a}\cdot k_b \,\tau_{(a,b)}, \nonumber \\
& d\mu_n  \equiv \frac{1}{2^2}
 \frac{d\L}{\L} \times \prod_{a=1}^n \frac{y_a \, dy_a}{C_a} \times  
\prod^{n-1}_{d=4} \,\frac{d\s_{d}}{ S^\tau_d}, \, \nonumber\\
& {\rm PT}^\tau(\a)  \equiv \tau_{(\a_1,\a_2)}\,\tau_{(\a_2,\a_3)}\cdots \tau_{(\a_n,\a_1)}, \,\, \nonumber\\
& \Delta^\tau_{(pqr)}  \equiv  \left( \tau_{(p,q)}\tau_{(q,r)}\tau_{(r,p)}\right)^{-1} \, , \nonumber\\
& \Delta^\tau_{(pqr|m)}  \equiv 
\s_p\Delta^\tau_{(qrm)}
- \s_m \Delta^\tau_{(pqr)} +\s_r  \Delta^\tau_{(mpq)} - \s_q  \Delta^\tau_{(rmp)}. ~~
\end{align} 
The  $\Gamma$  contour is defined by the $2n-3$ equations
\begin{equation}
\begin{cases}\L=0 \\ S^{\tau}_d=0\end{cases}\!\!\!{\rm for}~ d\neq \{p,q,r,m\}, ~C_{1}=0,\ldots, C_{ n} =0.
\end{equation}

As a simple example, let us consider the four-point amplitude, $A_4^{\phi^3}(1,2,3,4)$, with the gauge fixing $(pqr|m)=(412|3)$. 
By performing the integration, $\prod_a y_a, dy_a / C_a$, we first focus on the configuration in which the puncture pairs $\{\sigma_1, \sigma_2\}$ and $\{\sigma_3, \sigma_4\}$ lie on the upper and lower sheets of the curve, respectively,
\begin{align}
(y_1=+\sqrt{\s_1^2-\L^2},\s_1), \quad (y_2=+\sqrt{\s_2^2-\L^2},\s_2),  \\[-3pt]
(y_3=-\sqrt{\s_3^2-\L^2},\s_3),  \quad (y_4=-\sqrt{\s_4^2-\L^2},\s_4). \nonumber
\end{align} \\[-10pt]
Expanding  all elements of $A^{\phi^3}_4(1,2,3,4)$ around $\L=0$,  we obtain, to leading order,
\begin{align}
	 {\rm PT}^\tau(1,2,3,4)\Big|^{1,2}_{3,4}= \frac{\L^2}{2^2} \frac{1}{(\s_{12} \s_{2\zeta} \s_{\zeta 1} )} 
	\frac{1} { ( \s_{\zeta' 3} \s_{34} \s_{4\zeta'} )}, \nonumber
\end{align}
\begin{align}\label{TTeq}
&\frac{\Delta_{(412)} \Delta_{(412|3)}}{ S^{\tau}_3}
\Big|^{1,2}_{3, 4}  
=  
\frac{2^5}{\L^4} (\s_{\zeta 1} \, \s_{12}\, \s_{2\zeta})^2 \times
 \left( \frac{ 1 }{ s_{34}} 
\right) 
\nonumber  \\
&\hskip2.8cm  
\times (\s_{4\zeta'} \, \s_{\zeta' 3} \, \s_{34} )^2,
\end{align}
where two new fixed punctures arise at $\s_{\zeta}=\s_{\zeta'}=0$.
From the measure, $d\mu_4 = \frac{1}{2^2}\frac{d\L}{\L}$, we compute the $\L$ integral and the amplitude becomes,
\begin{align}
& 
A^{\phi^3}_4(1,2,3,4) \Big|^{1,2}_{3,4} = 
\frac{1}{2}  \frac{{\cal J}^{\phi^3}_3(1,2,\zeta)
\times {\cal J}^{\phi^3}_3 (\zeta',3,4)  }{s_{34}},
\end{align}
\noindent 
which has been factorized into a product of two amputated currents in the single-cover formulation, each carrying one off-shell leg,
$k_{\zeta}=k_3+k_4$ and $k_{\zeta'}=k_1+k_2$.
The overall factor $1/2$ cancels out after summing over mirrored configurations, {\it i.e.}, 
\begin{align}\label{schannel}
&
A^{\phi^3}_4(1,2,3,4) \Big|^{1,2}_{3,4} +A^{\phi^3}_4(1,2,3,4) \Big|_{1,2}^{3,4} \nonumber \\
&
=  
 \frac{{\cal J}^{\phi^3}_3(1,2,\zeta)
\times {\cal J}^{\phi^3}_3 (\zeta',3,4)  }{s_{34}}.
\end{align}
\noindent 
In a similar way, the factorization expansion,  $A^{\phi^3}_4(1,2,3,4) \Big|^{4,1}_{2,3} $, turns into, 
\begin{align}\label{tchannel}
&
A^{\phi^3}_4(1,2,3,4) \Big|^{4,1}_{2,3} +A^{\phi^3}_4(1,2,3,4) \Big|^{2,3}_{4,1} \nonumber \\
&
=  \frac{{\cal J}^{\phi^3}_3(2,3,\kappa)
\times {\cal J}_3^{\phi^3}(1,\kappa',4)  }{s_{23}} .
\end{align}
\noindent 
with $k_{\kappa}=k_4+k_1$ and $k_{\kappa'}=k_2+k_3$.

One can easily verify that the remaining configurations, for instance $ A^{\phi^3}_4(1,2,3,4)\big|^{1,3}_{2,4}$, $A^{\phi^3}_4(1,2,3,4)\big|^{1,2,3}_{4} $, and  $A^{\phi^3}_4(1,2,3,4)\big|^{1,2,3,4}$, vanish identically upon expansion and integration over $\Lambda$.

Therefore, the double-cover approach give us the four-point factorization relation
\begin{align}\label{f-pFactorized}
A^{\phi^3}_4(1,2,3,4) & =    \frac{{\cal J}^{\phi^3}_3(1,2,\zeta)
\times {\cal J}^{\phi^3}_3 (\zeta',3,4)  }{s_{34}}   \nonumber \\
&
+
\frac{{\cal J}^{\phi^3}_3(2,3,\kappa)
\times {\cal J}_3^{\phi^3}(1,\kappa',4)  }{s_{23}}\nonumber\\
&=\frac{1}{s_{34}}+\frac{1}{s_{23}},
\end{align}
\noindent 
where we have used the identity,  ${\cal J}^{\phi^3}_3(a,b,c)=1$. Fig.~\ref{Fig1} provides a schematic representation of the two factorization channels contributing to this four-point amplitude.

This factorization formula is easily obtained from the integration rules formulated by the author in \cite{Gomez:2016bmv}.

The analysis and computations carried out for the four-point example can be straightforwardly generalized to higher multiplicity.
Accordingly, Fig.~\ref{Fig1} summarizes all possible non-vanishing factorization contributions up to the eight-point level, corresponding to the gauge fixing $(pqr|m) = (n12|3)$.
{\small
\begin{figure}[h]
\centering
\hspace{-6.9cm}
\parbox[c]{5.0em}{\includegraphics[scale=0.20]{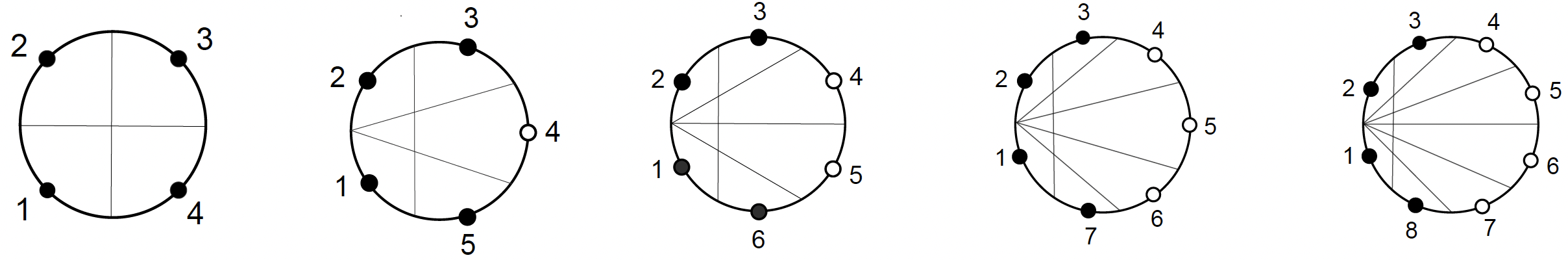}}
\caption{All factorization contributions for the four-, five-, six-, seven-, and eight-point. The white vertices indicate the remaining integrations over the $\sigma_i$ coordinates, taken around the corresponding scattering equations $S_i = 0$.}\label{Fig1} 
\end{figure}
}

Thus, for the setup $(pqr|m) = (n12|3)$, we obtain the following off-shell factorization formula:
\begin{align}\label{BAnpoint}
A^{\phi^3}_n(1,\ldots, n ) &=  
\frac{{\cal J}^{\phi^3}_{3}(1,2,\zeta)\times {\cal J}^{\phi^3}_{n-1}(\zeta',3,4,\ldots, n ) }{ s_{34\cdots n}}+
\nonumber\\
& 
\sum_{i=3}^{n-1}   \frac{ {\cal J}^{\phi^3}_{i}(2,3, \Omega_i, \kappa_i) \times {\cal J}^{\phi^3}_{n+2-i}(1,\kappa'_i, \Omega^\prime_i, n)}{s_{23\cdots i}} ,
\end{align}
where the off-shell legs are given by,
\begin{align}\label{offshelllegs}
&
k_\zeta=k_3+\cdots + k_n, \qquad\qquad\,\,\,	k_{\zeta'}=k_1+ k_2, \nonumber\\
&k_{\kappa_i}=k_{i+1}+\cdots + k_n+k_1, \quad 	k_{\kappa'_i}=k_2+\cdots + k_i.
\end{align}
The ordered sets, $\Omega_i$ and $\Omega'_i$, are defined as,
\begin{align}\label{omegasets}
\left.
\begin{matrix}
\hspace{-1.9cm}
\Omega_i=\{4,5,\ldots,i \}	\\
\Omega'_i=\{i+1,i+2,\ldots, n-1 \} \,\,\
\end{matrix}
\right\}\, i=4,\ldots, n-2,
\end{align}
and,
\begin{equation}
\Omega_3=\Omega'_{n-1}=\emptyset, \quad \Omega_{n-1}=\Omega'_{3}=\{4,5,\ldots, n-1 \}.
\end{equation}
Notice that, $\Omega_i \cup \Omega'_i = \{1,2,3,\ldots,n\} \setminus \{ p,q,r,m\}$, and $k_{\kappa_i}+k_{\kappa'_i}=0$.
Finally, we adopt the standard shorthand notation,
\begin{align}
 k_{i_1\cdots i_p} = k_{i_1}+\cdots +k_{i_p},\quad 	
 s_{i_1\cdots i_p} = \left( k_{i_1\cdots i_p}\right)^2. 	
\end{align}

It is useful to recall that the amputated currents \( \mathcal{J}^{\phi^3}(\alpha) \) are evaluated according to the prescription introduced by Naculich in Ref. \cite{Naculich:2015zha}. This means, the scattering equations are modified by the following kinematic shift:
\begin{equation}\label{eq:MSE}
S_a = \sum_{b \neq a}
\frac{s_{ab} + \Delta_{ab}}{\sigma_{ab}},
\end{equation}
where the only non-vanishing components of \( \Delta_{ab} \) are
\begin{align}\label{eq:DeltaMSE} 
&
\Delta_{cd} =k_c^2+k_d^2-k_\chi^2, \qquad
\Delta_{\chi c} =k_\chi^2+k_c^2-k_d^2, \nonumber\\
&
\Delta_{d\chi} = k_d^2+ k_\chi^2-k_c^2,
\end{align}
with \( c,d \in \{n,1,2,3\} \) and \( \chi \in \{\zeta, \zeta', \kappa_i, \kappa'_i\} \). In this case, it follows immediately that, $k_c^2=k_d^2=0$.

Making use of the ${\rm SL}(2,\mathbb{C})$ symmetry, the three scattering equations corresponding to the off-shell legs can be removed, \ie $\{c,d,\chi\}$. Consequently, the gauge-fixed integration measure retains the same form as in the fully massless case. Thus, for $n>4$, we can again use the factorization formula derived in Eq. \eqref{BAnpoint}, leading to,
\begin{align*}
& 
{\cal J}^{\phi^3}_{n-1}(\zeta',3,\ldots,n ) =    
 \frac{{\cal J}^{\phi^3}_{3}(\zeta',3,\tilde\zeta)  {\cal J}^{\phi^3}_{n-2}(\tilde\zeta',4,\ldots, n)}{s_{4\cdots n}}
\nonumber\\
&
+
\sum_{j=3}^{n-2}   \frac{ 
{\cal J}^{\phi^3}_{j}(3,4,\Omega_j,\kappa_j)
{\cal J}^{\phi^3}_{n+1-j}(\zeta',\kappa_j',\Omega_j',n)  }{s_{34\cdots j+1}}, 
\nonumber\\
&
{\cal J}^{\phi^3}_p(2,3,\Omega_p, \kappa_p) =    
 \frac{{\cal J}^{\phi^3}_{3}(2,3, \zeta) {\cal J}^{\phi^3}_{p-1}(\zeta',4,\ldots,\kappa_p)}{s_{45\cdots n1 }-
 k_{\kappa_p}^2
 }
\nonumber\\
&  
+
\sum_{j=3}^{p-1}  \frac{
{\cal J}^{\phi^3}_{j} (3,4, \tilde\Omega_j  , \tilde\kappa_{j})
 {\cal J}^{\phi^3}_{p+2-j}(2,\tilde\kappa'_{j}, \tilde\Omega'_j , \kappa_{p}) }{  s_{34\cdots j+1}}, 
\end{align*}
\begin{widetext}
\begin{align}\label{Crecursion}
{\cal J}^{\phi^3}_{n+2-q}(1,\kappa'_q,\Omega_q',n) =    
 \frac{{\cal J}^{\phi^3}_{3}(1,\kappa'_q,\zeta) {\cal J}^{\phi^3}_{n+1-q}(\zeta',q+1,\ldots, n)}{s_{q+1\cdots n}}
+\sum_{j=3}^{n+1-q}   \frac{
{\cal J}^{\phi^3}_{j}(\kappa'_q,q+1,\tilde\Omega_j, \tilde\kappa_j )
{\cal J}^{\phi^3}_{n+4-q-j}(1,\tilde\kappa'_j, \tilde\Omega'_j, n)} { s_{2 \cdots q+j-2}-
k_{\kappa'_q}^2
} , 
\end{align}
\end{widetext}
where $p = 4, \ldots, n-1$ and $q=3, \ldots, n-2$. The new labels and sets, namely $\{\tilde\zeta,\tilde\zeta',\tilde\kappa_j,\tilde\kappa'_j\}$, $\tilde\Omega_j$ and $\tilde\Omega'_j$, are defined in a similar way as in Eqs. \eqref{offshelllegs} and \eqref{omegasets}.

Together with Eq.~\eqref{BAnpoint}, the relations in Eq.~\eqref{Crecursion} furnish a recursive representation of the \(n\)-point amplitude.

\subsection{Comparison with Berends-Giele recursion}\label{ComparingBG}

Before proceeding, it is instructive to compare our construction with the standard Berends-Giele (BG) recursion, which provides a well established framework for computing amplitudes from off-shell currents \cite{Berends:1987me}.

Notice that the off-shell recursive relations obtained in Eqs.~\eqref{BAnpoint} and \eqref{Crecursion} for the $\mathrm{Tr}(\phi^3)$ theory differ from those derived using the BG (perturbiner) method by Mafra in \cite{Mafra:2016ltu}. In particular, while both approaches aim to construct multiparticle objects recursively, the structure and organization of the recursion in our case are distinct. 

For instance, let us consider the $n$-point scattering amplitude in the  $\mathrm{Tr}(\phi^3)$ theory within the BG framework, which is given by,
\begin{align}\label{BGprescription}
A^{\phi^3}_n(1,2,\ldots, n )=\lim_{k_n^2\to 0} 
s_{12\cdots n-1}\,{\cal J}^{\rm BG}_{n-1}(1,2,\ldots,n-1)\, \phi_n\, ,
\end{align}
where ${\cal J}_{n-1}^{\rm BG}$ denotes the BG current with one off-shell leg, defined recursively as,
\begin{align}
s_{i_1\cdots i_m}\,&{\cal J}^{\rm BG}_{m}(i_1,i_2,\ldots,i_m)\nonumber\\
&
= \sum_{a=1}^{m-1} {\cal J}^{\rm BG}_{a}(i_1,\ldots,i_a)\, {\cal J}^{\rm BG}_{m-a}(i_{a+1},\ldots,i_{m}),
\end{align}
with the boundary condition,
\begin{equation}
{\cal J}^{\rm BG}_{1}(i)=\phi_i=1.
\end{equation}

This is straightforward to see that the number of terms in Eq.~\eqref{BGprescription} is governed by the Catalan numbers, reflecting its direct correspondence with the standard Feynman diagram expansion. In this sense, the perturbiner approach naturally organizes the amplitude in terms of planar cubic graphs.

As a simple illustration, one can explicitly verify the following low-point currents,
\begin{align}
s_{12}\,{\cal J}^{\rm BG}_{2}(1,2)
&=
1\, , \\
s_{123}\,{\cal J}^{\rm BG}_{3}(1,2,3)
&=
{\cal J}^{\rm BG}_{2}(2,3)
+
{\cal J}^{\rm BG}_{2}(1,2)
\nonumber\\
&
=
\frac{1}{s_{23}}
+
\frac{1}{s_{12}}
\, ,
\\
s_{1234}\,{\cal J}^{\rm BG}_{4}(1,2,3,4)
&=
{\cal J}^{\rm BG}_{3}(2,3,4)
+
{\cal J}^{\rm BG}_{2}(1,2){\cal J}^{\rm BG}_{2}(3,4) \nonumber\\
&
+{\cal J}^{\rm BG}_{3}(1,2,3)
\nonumber\\
&=
\frac{1}{s_{234}}\left(
\frac{1}{s_{23}}
+
\frac{1}{s_{34}}
\right)
+
\frac{1}{s_{12}s_{34}}
\nonumber\\
&+
\frac{1}{s_{123}}\left(
\frac{1}{s_{12}}
+
\frac{1}{s_{23}}
\right)
\, .
\end{align}

Therefore, using the above currents and based on the prescription given in Eq.~\eqref{BGprescription}, it is straightforward to see that,
\begin{align}\label{fivepointBG}
A^{\phi^3}_3(1,2,3 )
&=
\lim_{k_3^2\to 0} 
s_{12}\,{\cal J}^{\rm BG}_{2}(1,2)\, \phi_3 
=
1\, , 
\nonumber\\
A^{\phi^3}_4(1,2,3,4)&
=
\lim_{k_4^2\to 0} 
s_{123}\,{\cal J}^{\rm BG}_{3}(1,2,3)\, \phi_4
\nonumber\\
&
=
\frac{1}{s_{23}}
+
\frac{1}{s_{12}}
\, ,
\nonumber\\
A^{\phi^3}_5(1,2,3,4,5)&
=
\lim_{k_5^2\to 0} 
s_{1234}\,{\cal J}^{\rm BG}_{4}(1,2,3,4)\, \phi_5
\nonumber\\
&=
\frac{1}{s_{234}}\left(
\frac{1}{s_{23}}
+
\frac{1}{s_{34}}
\right)
+
\frac{1}{s_{12}s_{34}}
\nonumber\\
&+
\frac{1}{s_{123}}\left(
\frac{1}{s_{12}}
+
\frac{1}{s_{23}}
\right)
\, ,
\end{align}
which precisely match the expected amplitudes.

On the other hand, within the CHY framework it is well known that  $A^{\phi^3}_3(1,2,3)=1$, while the four-point amplitude $A^{\phi^3}_4(1,2,3,4)$  was computed in Eq.~\eqref{f-pFactorized}. Now, using Eq.~\eqref{BAnpoint}, the five-point amplitude $A^{\phi^3}_5$ becomes, 
\begin{align}\label{fivepointchy}
&
A^{\phi^3}_5(1,2,3,4,5)  =  \frac{
 {\cal J}^{\phi^3}_4 (\zeta',3,4,5)  }{s_{345}}  + 
 \frac{{\cal J}_4^{\phi^3}(1,\kappa_3',4,5)  }{s_{23}}
 \nonumber \\
&
\qquad\qquad\qquad\quad
+
\frac{{\cal J}^{\phi^3}_4(2,3,4,\kappa_4)
 }{s_{234}}
\nonumber\\
&=\frac{1}{s_{12}} \left( \frac{1}{s_{34}}+\frac{1}{s_{123}} \right)
+\frac{1}{s_{23}} \left[ \frac{1}{s_{123}}+\frac{1}{s_{234} \left(1-\frac{s_{23}}{s_{234}}\right)} \right] \nonumber\\
&
+\frac{1}{s_{234}} \left[ \frac{1}{s_{34}}+\frac{1}{s_{23} \left(1-\frac{s_{234}}{s_{23}}\right)     } \right] 
,
\end{align}
where we have used ${\cal J}^{\phi^3}_3=1$. 
The amputated four-point currents ${\cal J}^{\phi^3}_4$ are obtained from Eq.~\eqref{Crecursion}, namely,
\begin{align}\label{currentsfive}
 &
{\cal J}^{\phi^3}_4(\zeta',3,4,5)=
\frac{1}{s_{123}} + \frac{1}{s_{34}}, \nonumber\\
&
{\cal J}^{\phi^3}_4(1,\kappa'_3,4,5)=
 \frac{1}{ s_{123}} +
 \frac{1}{s_{234}-s_{23}} ,\nonumber\\
 &
{\cal J}^{\phi^3}_4(2,3,4,\kappa_{4}) = 
\frac{1}{s_{23}-s_{234}}+
\frac{1}{s_{34}}.
\end{align}
 
By means of the simple identity,
\begin{align}
 \frac{1}{s_{23} s_{234}}\left[ \frac{1}{1-\frac{s_{234}}{s_{23}}}+\frac{1}{1-\frac{s_{23}}{s_{234}}}
		\right]=\frac{1}{s_{23} s_{234}},
\end{align}
clearly, the result in Eq.~\eqref{fivepointchy} is in agreement with that in Eq.~\eqref{fivepointBG}.

In contrast, the off-shell recursive relations obtained in Eqs.~\eqref{BAnpoint} and \eqref{Crecursion} for the $\mathrm{Tr}(\phi^3)$ theory differ significantly from the BG construction. Our prescription is based on a factorization-like approach, which leads to a different organization of the amplitude. This difference is already manifest at five-point: while Eq.~\eqref{fivepointBG} contains five terms, in agreement with the Catalan counting, Eq.~\eqref{fivepointchy} involves six terms. This overcounting signals that our representation is not minimal and, consequently, introduces spurious poles, as can be explicitly seen in Eq.~\eqref{fivepointchy}. Despite these structural differences, the final result remains consistent. 

Finally, it is worth emphasizing that the amputated currents arising in our approach differ from the standard BG currents. In particular, our construction allows for currents with up to three off-shell legs, whereas the BG currents involve only a single off-shell leg. Moreover, our amputated currents may contain spurious poles, while the BG currents exhibit only physical poles. This highlights a fundamental difference between the two approaches: while the perturbiner method provides a minimal and diagrammatic representation, our construction leads to a more redundant but factorization-driven description.

Although we have focused here on the $\mathrm{Tr}(\phi^3)$ theory, similar features can be observed in pure Yang-Mills.

\section{New On-Shell recursion for BAS}\label{SBArecursiveon}

The primary objective of this section is to determine a suitable basis of kinematic variables that permits the reconstruction of amputated currents from their on-shell counterparts, thereby enabling a smooth extension to Yang-Mills theory (YM).

We begin by analyzing the pole structure of the amputated currents derived in Eq.~\eqref{BAnpoint}, \emph{i.e.}, the poles appearing in Eq.~\eqref{Crecursion}.
Using momentum conservation, we eliminate the momentum \( k_n \), after which the poles take the form,
\begin{align}\label{commonsetMv}
s_{4\cdots n}&=-2k_{123}\cdot k_{4\cdots n-1}-(k_{4\cdots n-1})^2\nonumber\\
&=-\sum_{i=4}^{n-1}(s_{1i}+s_{2i}+s_{3i})-\sum_{i,j=4 \atop i<j}^{n-1}s_{ij}, \nonumber\\
s_{34\cdots j+1}&=\sum_{i,j=3 \atop i<j}^{j+1}s_{ij}, \quad j=3,\ldots, n-2, \nonumber\\
s_{4\cdots n1 }-k_{\kappa_p}^2 &= s_{23}-s_{2\cdots p}=s_{23}-\sum_{i,j=2 \atop i<j}^{p}s_{ij}
\nonumber\\
s_{q+1\cdots n } 
&=-2k_{1\cdots q}\cdot k_{q+1\cdots n-1} - (k_{q+1\cdots n-1} )^2 \nonumber\\
& =- \sum_{i=1}^q \sum_{j=q+1}^{n-1}  s_{ij}- \sum_{i,j=q+1 \atop i<j}^{n-1} s_{ij}, \nonumber\\
\nonumber\\
s_{2 \cdots q+j-2}-
k_{\kappa'_q}^2&=s_{2 \cdots q+j-2}-s_{2 \cdots q}=\sum_{i=1}^q \sum_{j=q+1}^{n-1}  s_{ij}
\end{align}

Clearly, given that $p = 4, \ldots, n-1$ and $q=3, \ldots, n-2$,
the amputated currents in Eq. \eqref{Crecursion} depend only on the $\frac{n(n-3)}{2} - 2$ \mand\ variables highlighted by the blue shading in the Gram matrix $s_{ij}$ in Eq. \eqref{Matrixnp}.
 We denote this set of kinematic variables as $\tilde{\mathbf{K}}_n$, for $n>4$, referring to it as the \textit{common set}.
{\small
\begin{equation}\label{Matrixnp}
\begin{tikzpicture}[scale=0.95,transform shape]
\matrix [matrix of math nodes,left delimiter=(,right delimiter=)] (m){
           0 & s_{12} &  s_{13}  &  s_{14} & \cdots &\,\, s_{1n-1}\,\,\,\,\,& s_{1n} \\
           {} & 0 & s_{23} & s_{24} & \cdots & s_{2n-1}\, & s_{2n} \\ 
            {} & {} & 0 & s_{34} & \cdots & s_{3n-1} \,& s_{3n}\\ 
            {} & {} & {} & {}  & \vdots \,\,\,\,\, & \vdots  & \vdots \\ 
            {} & {} & {} & {} & 0 & s_{n-2  n-1} & s_{n-2 \, n} \\ 
            {} & {} & {} & {} & {} & 0 & s_{n-1 \, n} \\
            {} & {} & {} & {} & {} & {} & 0 \\};
        \draw (m-1-3.north west) -- (m-2-3.north west)[opacity=1,thick];
         \draw (m-2-3.north west) -- (m-2-4.north west)[opacity=1,thick];
        \draw (m-1-3.north west) -- (m-1-6.north east)[opacity=1,thick];
         \draw (m-1-6.north east) -- (m-5-6.south east)[opacity=1,thick];
           \draw (m-2-4.north west) -- (m-3-4.south west)[opacity=1,thick];
            \draw (m-3-4.south west) -- (m-4-5.north west)[opacity=1,thick];
              \draw  (m-4-5.north west) -- (m-4-5.south west)[opacity=1,thick];
               \draw (m-4-5.south east) -- (m-5-6.south west)[opacity=1,thick];
               \draw (m-5-6.south west) -- (m-5-6.south east)[opacity=1,thick];
                  \draw (m-4-5.south west) -- (m-4-5.south east)[opacity=1,thick];
        \draw[ fill=blue, opacity=0.13] (m-4-5.south west) rectangle (m-1-5.north east);
        \draw[ fill=blue, opacity=0.13] (m-3-4.south west) rectangle (m-1-4.north east);
         \draw[fill=blue, opacity=0.13] (m-5-6.south west) rectangle (m-1-6.north east);
        \end{tikzpicture}\, .
        \end{equation}
        }\noindent 
A notable property of the common set is that the kinematic variables corresponding to the effective ``mass'' of the off-shell legs, \eg $k_{\kappa_i}^2=k_{2\cdots i}^2=m^2_i$, are not included in it, namely $ s_{12}\notin \tilde{\mathbf{K}}_n$ and $ s_{2\cdots i}\notin \tilde{\mathbf{K}}_n$, for $i=3,\ldots, n-1$. 

On the other hand, observe that \(\tilde{\mathbf{K}}_4 \) is not defined. This follows from the fact that \( {\cal J}^{\phi^3}_3 = 1 \), and consequently Eq. \eqref{Crecursion} does not apply in this case. Since the three-point current has no pole structure, the recursive construction becomes degenerate at four points.

To maintain a uniform treatment across multiplicities, and in particular to prepare for the pure YM extension, it is convenient to introduce a common kinematic basis also for \( n = 4 \). 
In defining the set \(\tilde{\mathbf{K}}_4\), we make use of the structural property of the common set discussed above.
From Eq.~\eqref{f-pFactorized}, one has,
\begin{align}\label{BD4p}
&
{A}^{\phi^3}_4(1,2,3,4)= 
\nonumber\\
&
\frac{{\cal J}^{\phi^3}_3(1,2,\zeta) {\cal J}^{\phi^3}_3(\zeta',3,4)}{s_{34}}
+
\frac{{\cal J}^{\phi^3}_3(2,3,\kappa) {\cal J}^{\phi^3}_3(1,\kappa',4) }{s_{23}} 
 \nonumber\\
&\!= \! 
\frac{1}{s_{34}}+ \frac{1}{s_{23}} .
\end{align}
Evidently, the effective ``mass'' of the off-shell legs in the amputated currents corresponds to the kinematic invariants, \( k_{\zeta}^2 = s_{34} \) and \( k_{\kappa}^2 = s_{23} \). Since these Mandelstam invariants can not belong to the common set, and at four points there are only three independent kinematic variables satisfying (with, $s_{13}+s_{23}+s_{34}=0$), we define the common set for $n=4$ as, $\{s_{13}\}$.

We therefore define the {\it the full common set} as,
\begin{equation}
\mathbf{K}_n \equiv \tilde{\mathbf{K}}_n \cup \{ s_{13} \}, \qquad n \geq 4,
\end{equation}
which contains \( \frac{n(n-3)}{2} - 1 \) independent Mandelstam variables.
This set corresponds precisely to the kinematic invariants selected by the polygon in matrix \eqref{Matrixnp}.

For example, for $n=5$, the amputated currents become (see Eq. \eqref{currentsfive}),
\begin{align}\label{BA5pointGM}
 &
{\cal J}^{\phi^3}_4(\zeta',3,4,5)=
\frac{1}{-s_{14}-s_{24}-s_{34}} + \frac{1}{s_{34}}, \nonumber\\
&
{\cal J}^{\phi^3}_4(1,\kappa'_3,4,5)=
 \frac{1}{ -s_{14}-s_{24}-s_{34}} +
 \frac{1}{s_{24}+s_{34}} ,\nonumber\\
 &
{\cal J}^{\phi^3}_4(2,3,4,\kappa_{4}) = 
\frac{1}{-s_{24}-s_{34}}+
\frac{1}{s_{34}},
\end{align}
where Eqs.~\eqref{Crecursion} and \eqref{commonsetMv} have been used.
One readily verifies that every Mandelstam invariant entering Eq.~\eqref{BA5pointGM} is contained in the full common set, $\mathbf{K}_5=\{s_{13},s_{14},s_{24},s_{34}\}$.

Let us now focus on a single off-shell contribution, \eg\ ${\cal J}^{\phi^3}_{n+2-i}\times {\cal J}^{\phi^3}_{i}$. As $\mathbf{K}_n$ is sufficient  to describe it, and the total number of independent kinematic variables for an $n$-point amplitude is $\frac{n(n-3)}{2}$, there remains one degree of freedom. 
This residual degree of freedom can naturally be assigned to the \mand\ variable corresponding to the effective ``mass'' of the off-shell leg, \ie $\{s_{2:i}\}$ in the present case. 
Given that ${\cal J}^{\phi^3}_{n+2-i}\times {\cal J}^{\phi^3}_{i}$ is independent of this kinematic invariant, we can consistently set it to zero (effectively an on-shell projection). This procedure enables the reconstruction of amputated currents directly from their on-shell counterparts.

For instance, in the five-point example discussed in Eq. \eqref{BA5pointGM}, we can choose the following sets of $\frac{5(5-3)}{2}=5$ independent kinematic variables, $\mathbf{K}_5 \cup \{s_{15}\}$ for ${\cal J}^{\phi^3}_4(2,3,4,\kappa_4)$, $\mathbf{K}_5 \cup \{s_{23}\}$ for ${\cal J}^{\phi^3}_4(1,\kappa'_3,4,5)$, and $\mathbf{K}_5 \cup \{s_{12}\}$ for ${\cal J}^{\phi^3}_4(\zeta',3,4,5)$.
With these choices, let us compute ${\cal J}^{\phi^3}_4(2,3,4,\kappa_4)$ from its on-shell counterpart, \ie the amplitude  ${A}^{\phi^3}_4(2,3,4,\kappa_4)$. It follows that,
\begin{align}\label{BA5pointOn}
 {A}^{\phi^3}_4(2,3,4,\kappa_4)\Big|_{\mathbf{K}_5}&=
\left.\left[\frac{1}{(k_4+k_{\kappa_4})^2}+
\frac{1}{s_{34}}\right]\right|_{\mathbf{K}_5}\nonumber\\
&
=
\left.\left[\frac{1}{2 k_4 \cdot k_{\kappa_4}}+
\frac{1}{s_{34}}\right]\right|_{\mathbf{K}_5}\nonumber\\
&
=
\frac{1}{-s_{24}-s_{34}}+
\frac{1}{s_{34}},
\end{align}
where we have used $k_{\kappa_4}=k_5+k_1=-k_2-k_3-k_4$, together with the on-shell condition (projection) $k_{\kappa_4}^2=s_{15}=0$.
The result obtained in this way agrees precisely with Eq. \eqref{BA5pointGM}, and we therefore arrive at the identity,
\begin{equation}
{A}^{\phi^3}_4(2,3,4,\kappa_4)\Big|_{\mathbf{K}_5}={\cal J}^{\phi^3}_4(2,3,4,\kappa_4).
\end{equation}
The corresponding results for the other amputated currents follow in a completely analogous manner and can be checked directly, 
\begin{align}\label{a4k5}
{A}^{\phi^3}_4(1,\kappa'_3,4,5) \Big|_{\mathbf{K}_5}
&
  =
{\cal J}^{\phi^3}_4(1,\kappa'_3,4,5) ,
 \nonumber\\
{A}^{\phi^3}_4(\zeta',3,4,5)\Big|_{\mathbf{K}_5} 
&
={\cal J}^{\phi^3}_4(\zeta',3,4,5).
\end{align}

With all these ingredients in place, we are now in a position to promote the factorization relation in Eq.~\eqref{BAnpoint} to an on-shell recursive relation.
This leads to, 
\begin{align}\label{BAnpointon}
A^{\phi^3}_n(1,\ldots , n ) &=  
\frac{{A}^{\phi^3}_{3}(1,2,\zeta)\times {A}^{\phi^3}_{n-1}(\zeta',3,4,\ldots, n ) \Big|_{\mathbf{K}_n} }{ s_{12}}+
\nonumber\\
& 
\!\!\!\!\!\!\!\!\!
\sum_{i=3}^{n-1}   \frac{ {A}^{\phi^3}_{i}(2,3, \Omega_i, \kappa_i) \times {A}^{\phi^3}_{n+2-i}(1,\kappa'_i, \Omega^\prime_i, n) \Big|_{\mathbf{K}_n}}{s_{23\cdots i}} .
\end{align}

Although the above on-shell recursive relation was derived within the DC framework using the gauge choice \( (pqr|m) = (n12|3) \), one may adopt a different gauge fixing and obtain an alternative recursive representation. This generalization follows straightforwardly by applying the procedure outlined in the previous sections.

Additionally, some higher point examples are provided in \cite{mathematica}.

\subsection{Connection to BCFW On-Shell Recursion}\label{ComparingBCFW}

The Britto-Cachazo-Feng-Witten (BCFW) recursion \cite{Britto:2005fq,Arkani-Hamed:2008bsc} is a well known on-shell construction in which a pair of external momenta is deformed by a complex parameter, while preserving both momentum conservation and the on-shell conditions.

The recursion relation obtained from this method, which is likewise based on factorization, takes a form closely resembling that of Eq.~\eqref{BAnpointon}. A key difference between the BCFW framework and our approach, however, is that our construction avoids the need to analyze boundary contributions associated with large complex deformations. Nevertheless, it is possible to combine both formalisms to develop a more powerful framework.

To proceed, let us consider the following complex deformation of the external momenta:
\begin{align}
k_1(z) &= k_1 + z q \, , \nonumber \\
k_2(z) &= k_2 - z q \, ,
\end{align}
where $z$ is a complex parameter and $q$ is a null vector satisfying,
\begin{equation}
q^2 = k_1 \cdot q = k_2 \cdot q = 0 .
\end{equation}
Notice that under these conditions, the deformed momenta remain on-shell, $k_1^2(z)=k_2^2(z)=0$, and momentum conservation is preserved, $k_1(z) + k_2(z) + k_3 + \cdots + k_n = 0$.

Thus, using the deformed kinematic data $\{ k_1(z), k_2(z), k_3, \ldots, k_n \}$, the $n$-point amplitude $A^{\phi^3}_n$ can be written as,
\begin{align}
A^{\phi^3}_n(1,2,\ldots , n )= \oint_{|z|=\varepsilon}   \frac{A^{\phi^3}_n(1,\ldots , n )(z)}{z}\, dz \, .
\end{align}

Applying the global residue theorem, one readily finds that the amplitude $A^{\phi^3}_n(1,2,\ldots , n )$
can be expressed as, 
\begin{align}\label{BABCFW} 
A^{\phi^3}_n&(1,2,\ldots , n )\nonumber\\
&=
\sum_{i=3}^{n-1}   \frac{ {A}^{\phi^3}_{i}(2,3, \Omega_i, \kappa_i)(z_i) \times {A}^{\phi^3}_{n+2-i}(1,\kappa'_i, \Omega^\prime_i, n)(z_i) }{s_{23\cdots i}}
\nonumber\\
&-{\rm Res}_{z=\infty}\left[\frac{A^{\phi^3}_n(1,\ldots , n )(z)}{z}\right]\, ,
\end{align}
where $z_i$ is the solution to the pole condition, $s_{23\cdots i}(z)=(k_2(z)+k_3+\cdots+k_i)^2=0$, namely,
\begin{equation}
z_i =\frac{s_{23\cdots i}}{2 \, q\cdot \kappa'_{i}}=-\frac{s_{23\cdots i}}{2 \, q\cdot \kappa_{ i}}=\frac{s_{23\cdots i}}{2 \, q\cdot k_{23\cdots i}}.
\end{equation}

By comparing Eqs.~\eqref{BAnpointon} and \eqref{BABCFW}, one can combine both approaches to reveal the following identities:
\begin{align}
&\sum_{i=3}^{n-1}   \frac{ {A}^{\phi^3}_{i}(2,3, \Omega_i, \kappa_i)(z_i) \times {A}^{\phi^3}_{n+2-i}(1,\kappa'_i, \Omega^\prime_i, n)(z_i) }{s_{23\cdots i}}
\nonumber\\
&=
\sum_{i=3}^{n-1}   \frac{ {A}^{\phi^3}_{i}(2,3, \Omega_i, \kappa_i) \times {A}^{\phi^3}_{n+2-i}(1,\kappa'_i, \Omega^\prime_i, n) \Big|_{\mathbf{K}_n}}{s_{23\cdots i}} \, , \label{fidentity}
\\
&
-{\rm Res}_{z=\infty}\left[\frac{A^{\phi^3}_n(z)}{z}\right]
=
\frac{ {A}^{\phi^3}_{n-1}(\zeta',3,4,\ldots, n ) \Big|_{\mathbf{K}_n} }{ s_{12}}\, , \label{sidentity}
\end{align}
where we have used ${A}^{\phi^3}_{3}=1$. 
This is an important result, as it enables us to rewrite the boundary contribution of the BCFW method in a purely on-shell form. On the other hand, the first identity in Eq.~\eqref{fidentity} does not hold term by term. To clarify this point, we present a simple example and analyze the necessary conditions required to establish the identity at the level of individual terms.

Let us consider the first nontrivial example, namely the five-point amplitude $A^{\phi^3}_5(1,2,3,4,5)$. Making use of Eq.~\eqref{BABCFW}, {\it i.e.},
\begin{align}\label{BABCFW5}
A^{\phi^3}_5&(1,2,3,4, 5 ) =
\frac{A^{\phi^3}_4(1,\kappa'_{3},4,5)(z_3)}{s_{23}}
\nonumber\\
&
+\frac{ A^{\phi^3}_4(2,3,4,\kappa_{4})(z_4) }{s_{234}}
-{\rm Res}_{z=\infty}\left[\frac{A^{\phi^3}_5(z)}{z}\right]\, ,
\end{align}
and incorporating the boundary contribution given in Eq.~\eqref{sidentity}, more precisely,
\begin{align}
-{\rm Res}_{z=\infty}\left[\frac{A^{\phi^3}_5(z)}{z}\right]
=
\frac{ {A}^{\phi^3}_{4}(\zeta',3,4,5 ) \Big|_{\mathbf{K}_5} }{ s_{12}}\, , 
\end{align}
where ${A}^{\phi^3}_{4}(\zeta',3,4,5 ) \Big|_{\mathbf{K}_5}$ has been computed previously in Eqs. \eqref{BA5pointGM} and \eqref{a4k5}, one obtains,
\begin{align}\label{5pBCFW}
&
A^{\phi^3}_5(1,2,3,4, 5 )=
\frac{1}{s_{23}}\left[
\frac{1}{ s_{123}}
+
\frac{1}{s_{234}\left(1-\frac{s_{23}\,q\cdot k_5}{s_{234}\, q\cdot k_{45}}\right)} 
\right]+
\nonumber\\
&
\frac{1}{s_{234}} 
\left[
\frac{1}{s_{34}} + \frac{1}{s_{23}\left(1-\frac{s_{234}\,q\cdot k_{45}}{s_{23}\, q\cdot k_{5}}\right)}
\right]
+
\frac{1}{s_{12}}\left(
\frac{1}{s_{34}} + \frac{1}{s_{123}}
\right).
\end{align}
Clearly, using the identity,
\begin{align}\label{}
\frac{1}{s_{23},s_{234}}\left[
\frac{1}{1-\frac{s_{23}, q\cdot k_5}{s_{234} , q\cdot k_{45}}}
+
\frac{1}{1-\frac{s_{234} , q\cdot k_{45}}{s_{23} , q\cdot k_5}}
\right]
= \frac{1}{s_{23},s_{234}},
\end{align}
one immediately verifies that Eq.~\eqref{5pBCFW} reproduces the correct amplitude.

An interesting observation is the close similarity between Eqs.~\eqref{fivepointchy} and \eqref{5pBCFW}. In order to establish an exact correspondence, it is sufficient to impose the additional condition $k_4 \cdot q = 0$, which is consistent since $q$ is a complex vector. Under these conditions, namely
\begin{equation}
q^2 = k_1 \cdot q = k_2 \cdot q = k_4 \cdot q = 0,
\end{equation}
Eq.~\eqref{fidentity} becomes a term-by-term identity at five-point. In particular, one finds
\begin{align}\label{Id5point}
{A}^{\phi^3}_{4}(1,\kappa'_3, 4, 5)(z_3) &= {A}^{\phi^3}_{4}(1,\kappa'_3, 4, 5)\Big|_{\mathbf{K}_5}\, ,
\nonumber\\
{A}^{\phi^3}_{4}(2,3,4,\kappa_4)(z_4) &= {A}^{\phi^3}_{4}(2,3,4,\kappa_4)\Big|_{\mathbf{K}_5}\, .
\end{align}

A similar analysis can be extended to higher multiplicities and to Yang-Mills theory, where one expects analogous structures to emerge. We leave a detailed study of these aspects for future work \cite{inpreparation}.

\section{Pure Gluons}

 In the subsequent sections, we consider color-ordered Yang-Mills amplitudes, which we denote by \( A_n(1,\ldots,n) \).  Each label \( a \) specifies an external particle carrying momentum \( k_a \) and polarization vector \( \epsilon_a \).

Given a similar setup as in the BAS case: by choosing the gauge-fixing  \( (pqr|m) = (n12|3) \), and removing rows and columns $(1,3)$ from the Pfaffian. In \cite{Bjerrum-Bohr:2018lpz}, we demonstrated  that the $n$-point pure YM amplitude is factorized as,
\begin{align}\label{YMnpointOff}
&
A_n(1,\ldots,n) =  
 \frac{ \sum_M {\cal J}_{3}\big(\underline{1},2,\underline{\zeta}^{M}\big) \times {\cal J}_{n-1}\big(\underline{\zeta'}^M,\underline{3},\ldots, n\big) }{s_{12}}
\nonumber\\
&
+
\sum_{i=3}^{n-1}
\Bigg[
\frac{\sum_M {\cal J}_{i}\big({2},{\underline{3}},\Omega_i, {\underline{\kappa}}^M_{i} \big)
\times
{\cal J}_{n+2-i}\big({\underline{1}},{\underline{\kappa}}'^M_{i},\Omega'_i, {n}\big)
}{s_{23\cdots i}} 
\nonumber \\
&
+2 
\left.
\,\frac{\sum_L
{\cal J}_{i}({\underline{2}},{3},\Omega_i, {\underline{\kappa}}^L_{i})
\times
{\cal J}_{n+2-i}({\underline{1}},{\underline{\kappa}}'^L_{i},\Omega'_i, {n})
}{s_{23\cdots i}}\right|^{1\leftrightarrow 2}
\Bigg],
\end{align}
where \( \mathcal{J}_m \) denotes the Yang-Mills amputated currents with off-shell legs \( \{\zeta,\zeta',\kappa_i,\kappa_i'\} \) defined in Eq.~\eqref{offshelllegs}, 
while the sets \( \Omega_i \) and \( \Omega_i' \) are specified in Eq.~\eqref{omegasets}.
The underline labels indicate rows/columns removed from the Pfaffian.

In this context, $\sum_M$ represents the sum over all off-shell polarization states, whereas $\sum_L$ refers specifically to the longitudinal sector. Therefore, the associated completeness relations are  given by
\begin{equation}
\sum_M \epsilon^{M,\mu}_{I}  \epsilon^{M,\nu}_{J} = \eta^{\mu \nu} 
,\qquad
\sum_L \epsilon^{L,\mu}_{I}  \epsilon^{L,\nu}_{J} = \frac{k_I^{\mu}\,k_J^{\nu}}{k_I\cdot k_J},
\end{equation}
with $I\in\{\zeta,\kappa_i \}$ and $J\in\{\zeta',\kappa'_i \}$.
Finally, the superscript $(1 \leftrightarrow 2)$ means the labels 1 and 2 must be flipped.

In complete analogy with the BAS case, the YM amputated currents are evaluated using the modified scattering equations in Eq.~\eqref{eq:MSE}. Accordingly, the Pfaffian entries \( \frac{s_{ab}}{\sigma_{ab}} \) are deformed by the addition of the term \( \Delta_{ab} \), as defined in Eq.~\eqref{eq:DeltaMSE}, leading to the substitution (see Ref. \cite{Bjerrum-Bohr:2018lpz}),
\begin{equation}
\frac{s_{ab}}{\sigma_{ab}}\,\to\,
 \frac{s_{ab}+\Delta_{ab}}{\sigma_{ab}}.
\end{equation}

As shown in Refs.~\cite{Bjerrum-Bohr:2018lpz}, the amputated currents appearing in Eq.~\eqref{YMnpointOff} are not transverse.
Schematically, one finds,
\begin{equation}
	{\cal J}_m\Big|_{\eps_\kappa \to k_\kappa} \propto k_\kappa^2\, \neq \, 0,
\end{equation}
a property that plays a crucial role in reproducing the contact terms. 
 
Evidently, the this factorization formula mirrors the BAS structure in Eq. \eqref{BAnpoint}. We will show that the BAS results naturally extend to pure YM.

\section{On-shell Relation at lower Points}

As the following sections aim to show how the off-shell factorization formula in Eq. \eqref{YMnpointOff} naturally extends to an on-shell form, within the framework previously developed, we begin with illustrative examples.
 
 The simplest non-trivial case is the four-point amplitude $A_4(1,2,3,4)$. From Eq. \eqref{YMnpointOff}, this amplitude receives three off-shell contributions,
\begin{align}\label{YM4point} 
A_4(1,2,3,4)=& 
-\frac{ 
\sum_M {\cal J}_{3}({\underline{1}},{2},{\underline{\zeta}}^M)\times {\cal J}_{3}({\underline{\zeta}}'^M,{4},{\underline{3}})
}{s_{12}}
\nonumber\\
&
-  
\frac{\sum_M 
 {\cal J}_{3}({\underline{1}},4,{\underline{\kappa}}'^M) 
 \times
 {\cal J}_{3}\big({\underline{\kappa}^M}, {2},{\underline{3}} \big)
 }{s_{23}}
\nonumber\\
&
-2 \, 
\left.
\frac{\sum_L 
{\cal J}_{3}( {\underline{\kappa}}^L,{3},{\underline{2}})
\times
{\cal J}_{3}({\underline{\kappa}}'^L,{4},{\underline{1}}) 
}{s_{23}}\right|^{1\leftrightarrow 2},
\end{align}
with $k_\zeta=k_3+k_4=-k_{\zeta'}$ and $k_\kappa=k_4+k_1=-k_{\kappa'}$. We have made use of the cyclic and reflection identities,
\begin{align}
&A_n(i_1,i_2,\ldots, i_n)=A_n(i_2,\ldots, i_n,i_1)\nonumber\\
&A_n(i_1,i_2,\ldots, i_n)=(-1)^n	\,A_n(i_n,i_{n-1},\ldots ,i_1),
\end{align}
which can be straightforwardly extended to the amputated currents ${\cal J}_m$.

Here, our focus is solely on the transition to the on-shell case, guided by the results obtained previouly.

First, notice that the non-longitudinal contributions, specifically, the first two lines, exhibit the same structure as in the BAS case. Thus, building upon the framework developed in Section~\ref{SBArecursiveon}, we select the following sets of independent kinematic variables for the amputated currents and subsequently compute the associated on-shell contributions, namely,
\begin{align}
&1.\, \mathbf{K}_4 \cup \{s_{34}\} \,\,\,\,\, \text{ for }\quad
\sum_M {\cal J}_{3}({\underline{1}},{2},{\underline{\zeta}}^M)\times {\cal J}_{3}({\underline{\zeta}}'^M,{4},{\underline{3}}), 
\nonumber\\
&
\text{and the on-shell side, }
\sum_T 
{A}_{3}(1,{2},{\zeta}^T) \times {A}_{3}({\zeta}'^T,{4},{3})\Big|_{\mathbf{K}_4}, \nonumber\\
&
2.\, \mathbf{K}_4 \cup \{s_{23}\} \,\,\,\,\, \text{ for }\quad
\sum_M 
 {\cal J}_{3}({\underline{1}},4,{\underline{\kappa}}'^M) 
 \times
 {\cal J}_{3}\big({\underline{\kappa}^M}, {2},{\underline{3}} \big),
\nonumber\\
& 
\text{and the on-shell side, }  
\sum_T A_{3}(1,4,\kappa'^T)\times A_{3}(\kappa^T,2,3)\Big|_{\mathbf{K}_4} ,
 \nonumber
\end{align}
where $\mathbf{K}_4=\{s_{13}\}$. The sum $\sum_T$ runs over all possible on-shell (transverse) polarization vectors, which implies the completeness relation,
\begin{equation}\label{transverse}
\sum_T \epsilon^{T,\mu}_{I} \epsilon^{T,\nu}_{J} = \eta^{\mu \nu} - \frac{k_{I}^{\mu}  q^\nu - q^\mu  k_{J}^\nu}{k_{I}\cdot q},\quad k_I^2=0,
\end{equation}
with $I\in\{\zeta,\kappa \}$ and $J\in\{\zeta',\kappa' \}$.

Let us first to recall the expression for the three-point 
amputated current ${\cal J}_3 (\underline{1},2, \underline{\zeta}^M)$. From the CHY approach, it takes the form,
\begin{align}\label{YM123}
&
{\cal J}_3 (\underline{1},2, \underline{\zeta}^M)  
=
(\s_{12}\,\s_{2\zeta}\,\s_{\zeta 1})^2 \,\, {\rm PT}(1,2,\zeta)\nonumber\\
&
\qquad\,\,\,\qquad\,\,\,\,\,\, \times
\frac{1}{\s_{1\zeta}}
\,
{\rm Pf}\left[
{\small
\begin{matrix}
0 & - \frac{\eps_{1}\cdot k_{2}}{\s_{12}} & - \mathsf{C}_{22}   & -  \frac{\eps_{\zeta}^M\cdot  k_{2}}{\s_{\zeta 2}} \\
 \frac{\eps_{1 }\cdot   k_{2}}{\s_{12}} & 0 & \frac{\eps_{1}\cdot \eps_{2}}{\s_{12}} & \frac{\eps_{1}\cdot \eps_{\zeta}^M}{\s_{1\zeta}}\\
 \mathsf{C}_{22} &  \frac{\eps_{2}\cdot \eps_{1}}{\s_{21}} &  0 & \frac{\eps_{2}\cdot \eps_{\zeta}^M}{\s_{2\zeta}} \\
\frac{\eps_{\zeta}^M \cdot k_{2}}{\s_{\zeta 2}}  &   \frac{\eps_{\zeta}^M \cdot \eps_{1}}{\s_{\zeta 1}} &  \frac{\eps_{\zeta}^M\cdot \eps_{2}}{\s_{\zeta 2}} &  0  \\
\end{matrix}}
\right] 
\nonumber \\
& =
(\eps_1 \cdot \eps^M_\zeta)(\eps_2\cdot { k_\zeta})-
(\eps_2\cdot \eps^M_\zeta)(\eps_1\cdot { k_\zeta }) - (\eps_1\cdot \eps_2) (\eps^M_\zeta\cdot { k_2}) 
,\nonumber\\
\end{align}
with, $\mathsf{C}_{22}= -\frac{\eps_2 \cdot k_1}{\s_{21}}-\frac{\eps_2 \cdot k_\zeta}{\s_{2\zeta}}=(\eps_2 \cdot k_\zeta)\,\frac{\s_{\zeta 1}}{\s_{12}\, \s_{2 \zeta}} $. 
Employing the reflection identity together with an appropriate relabeling of the external legs, we can compute the other amputated currents. It leads to,
\begin{align}\label{}
&
\sum_M {\cal J}_{3}({\underline{1}},{2},{\underline{\zeta}}^M)\times {\cal J}_{3}({\underline{\zeta}}'^M,{4},{\underline{3}})
 = (n_{(124)}\cdot \epsilon_3),\nonumber\\
 &
 \sum_M 
 {\cal J}_{3}({\underline{1}},4,{\underline{\kappa}}'^M) 
 \times
 {\cal J}_{3}\big({\underline{\kappa}^M}, {2},{\underline{3}} \big)
 = (n_{(142)}\cdot \epsilon_3),
\end{align}
where we have used of the kinematic identity, $k_2\cdot k_4= 2^{-1} s_{13}$, and have introduced the current (see Refs.~\cite{Bjerrum-Bohr:2018lpz,Du:2017kpo}),
\begin{align}\label{currentn}
&
n^\mu_{(i_1\ldots i_n)}\equiv	(k_{i_1}\cdot\eps_{i_2})(k_{i_1 i_2}\cdot \eps_{i_3})\cdots (k_{i_1:i_{n-1}}\cdot \eps_{i_n})
\nonumber\\
&
\left( \eps_{i_1} \cdot
\exp\left[
\frac{-f_{i_2}}{(k_{i_1}\cdot \eps_{i_2})}
\right] \cdots 
\exp\left[
\frac{-f_{i_n}}{(k_{i_1:i_{n-1}}\cdot \eps_{i_n})}
\right]
\right)^\mu,
\end{align}
with $f^{\mu\nu}_i \equiv k^\mu_i \eps^\nu_i - k^\nu_i \eps^\mu_i$ and $(\eps_i\cdot \eps_i)=0$ (this is to guarantee that only linear contributions arise from the exponentials).

Turning to the on-shell side, we begin by  
considering the three-point YM amplitude,
\begin{align}\label{YM3point}
A_3(1,2,3)&=n_{(1,2)}\cdot \eps_{3}\nonumber\\
&\equiv N_{(1,2,3)}\, ,
\end{align}
where $N_{(1,2,3)}$ 
denotes the three-point BCJ numerator. 
Thus, it is straightforward to check,
\begin{align}\label{YM4On}
\sum_T A_{3}(1,2,\zeta^T)\times A_{3}(\zeta'^T,4,3)\Big|_{\mathbf{K}_4}  &= (n_{(124)}\cdot \epsilon_3), \nonumber\\
\sum_T A_{3}(1,4,\kappa'^T) \times A_{3}(\kappa^T,2,3)\Big|_{\mathbf{K}_4}  &=(n_{(142)}\cdot \epsilon_3),
\end{align}
which is in complete agreement with the expected result, \ie
\begin{align}\label{}
&
\sum_M {\cal J}_{3}({\underline{1}},{2},{\underline{\zeta}}^M) {\cal J}_{3}({\underline{\zeta}}'^M,{4},{\underline{3}})
\nonumber\\
&
\qquad\qquad\qquad\qquad
 = \sum_T A_{3}(1,2,\zeta^T) A_{3}(\zeta'^T,4,3)\Big|_{\mathbf{K}_4}
 ,\nonumber\\
 &
 \sum_M 
 {\cal J}_{3}({\underline{1}},4,{\underline{\kappa}}'^M) 
 {\cal J}_{3}\big({\underline{\kappa}^M}, {2},{\underline{3}} \big)
 \nonumber\\
 &
 \qquad\qquad\qquad\qquad
 = 
\sum_T A_{3}(1,4,\kappa'^T)  A_{3}(\kappa^T,2,3)\Big|_{\mathbf{K}_4} .
\end{align}

Now, we turn our attention to the longitudinal sector. From the last line in Eq. \eqref{YM4point}, we can see that $s_{13}$ is a spurious pole, which implies the numerator must be proportional to it.
Indeed, for this contribution, it is straightforward to observe that the \mand\ variable corresponding to the ``mass'' of the off-shell leg, namely $k_{13}^2 = s_{13}$, decouples from the current, which enables an on-shell formulation. To achieve this, we use the formula found by He \etal in \cite{Dong:2021qai} (see also \cite{Zhou:2022orv}),
\begin{align}\label{YMBAnp}
&
A_{n}(1,2,\ldots ,n)= 
\nonumber\\
&   
\sum_{\Xi} (\epsilon_n\cdot F_{\Xi^t} \cdot \epsilon_1)\,\,A^{{\rm YM}\oplus\phi^3}_{n}(1,\Xi,n|1,2,\ldots ,n),
\end{align}
where  $A^{{\rm YM}\oplus\phi^3}_n$ is the single-trace amplitude involving gluons and BAS interactions. The first ordering in the argument specifies the scalar sector, whereas the second one corresponds to the complete ordering of the external legs. 
The sum \( \sum_{\Xi} \) extends over all subsets 
\( \Xi \subseteq \{2,\ldots,n-1\} \), including the empty set. 
We regard \( \Xi \) as an ordered set, 
\( \Xi = \{\xi_1,\ldots,\xi_r\} \) with \( r \ge 0 \), and define,
\begin{equation}
	(\epsilon_n \cdot F_{\Xi^t} \cdot \epsilon_1)
	\equiv
	(\epsilon_n \cdot f_{\xi_r} \cdot f_{\xi_{r-1}} \cdots f_{\xi_1} \cdot \epsilon_1),
\end{equation}
where \( \Xi^t \) denotes the reversed ordering of \( \Xi \).

Thus, from Eq. \eqref{YMBAnp}, it is simple to verify that the longitudinal contributions can be recast as, 
\begin{align}\label{YM3OnL}
\frac{2}{k_\kappa^2}
 {\cal J}_{3}({\underline{\kappa}}^l,3,{\underline{2}}) &= (\epsilon_2\cdot \epsilon_3)\,A^{\phi^3}_{3}(\kappa,3,2| \kappa,3,2)\nonumber\\
 &= (\epsilon_2\cdot \epsilon_3), \nonumber\\
 \frac{2}{k_{\kappa'}^2}
 {\cal J}_{3}({\underline{\kappa}}'^l,4,{\underline{1}}) &= (\epsilon_1\cdot \epsilon_4)\,A^{\phi^3}_{3}(\kappa',4,1| \kappa',4,1)\nonumber\\
 &= (\epsilon_1\cdot \epsilon_4),
 \end{align}
where $k_\kappa^2 = k_{\kappa'}^2 = s_{23}$, and $\kappa^{l}$ ($\kappa'^{l}$) denotes the longitudinal projection obtained by setting $\epsilon_\kappa \to k_\kappa$ ($\epsilon_{\kappa'} \to k_{\kappa'}$). 

On the other hand, in order to formulate a recursive relation purely within Yang-Mills theory, the single-trace ${\rm YM} \oplus \phi^3$ amplitudes must first be rewritten entirely in terms of gluonic amplitudes. 
A systematic way to relate the $A^{{\rm YM} \oplus \phi^3}$ contributions to pure Yang-Mills amplitudes is through the use of \emph{transmutation operators}. As shown by Cheung \textit{et al.} in \cite{Cheung:2017ems}, this mapping takes the explicit form,
\begin{align}\label{YMpBA}
A^{{\rm YM} \oplus \phi^3}_n(\b|\gamma_1,\ldots,\gamma_n)	=
\mathcal{T}_{[\b]} \, A_n(\gamma_1,\ldots,\gamma_n),
\end{align}
where $\b = \{ \b_1, \b_2, \ldots, \b_m \}$, $m \leq n$, and the operator $\mathcal{T}_{[\b]}$ is defined as,
\begin{equation}
\mathcal{T}_{[\b ]}\equiv {\cal T}_{\b_1\b_m}\prod_{i=2}^{m-1}{\cal T}_{\b_{i-1}\b_{i}\b_m} ,	
\end{equation}
with $\mathcal{T}_{ij}\equiv \partial_{\epsilon_i\cdot \epsilon_j}$ and $\mathcal{T}_{ijr}\equiv \partial_{k_i\cdot \epsilon_j}-\partial_{k_r\cdot \epsilon_j}$.  
Hence, the longitudinal contributions obtained in Eq. \eqref{YM3OnL} become,
\begin{align}\label{YM3OnLA}
\frac{2}{k_{\kappa}^2}
 {\cal J}_{3}({\underline{\kappa}}^l,{i},{\underline{j}})&= (\epsilon_j\cdot \epsilon_i) {\cal T}_{[\kappa,i,j]}A_{3}(\kappa,i,j).
 \end{align}

Considering the results found in Eqs. \eqref{YM4On} and \eqref{YM3OnLA} into Eq.\eqref{YM4point}, we observe that,
\begin{align}\label{YM4pointOn} 
&
-\frac{\sum_T A_{3}(1,2,\zeta^T)A_{3}(\zeta'^T,4,3)\Big|_{\mathbf{K}_4} }{s_{12}}
\nonumber\\
&
- 
\frac{\sum_T A_{3}(1,4,\kappa'^T)A_{3}(\kappa^T,2,3)\Big|_{\mathbf{K}_4}
}{s_{23}} - \frac{1}{2}
\Big[(\epsilon_1\cdot \epsilon_4)(\epsilon_2\cdot \epsilon_3)
\nonumber\\
&
\quad\,
{\cal T}_{[\kappa',4,1]} {\cal T}_{[\kappa,3,2]}
A_{3}(1,\kappa',4)  A_{3}(2,3,\kappa)\Big]\Big|^{1\leftrightarrow2}
\nonumber\\
&
= 
\frac{-(n_{(124)}\cdot \epsilon_3) }{s_{34}}
+ 
\frac{-(n_{(142)}\cdot \epsilon_3)}{s_{23}}+\frac{1}{2}(\epsilon_1\cdot \epsilon_3)(\epsilon_2\cdot \epsilon_4)  \nonumber\\
&
=  A_4(1,2,3,4),
\end{align}
which is precisely the pure YM amplitude $A_4(1,2,3,4)$, confirming the validity of extending the off-shell factorization formula to its on-shell counterpart. 
 
We now turn to the five-point amplitude, which presents a more interesting and richer structure. 
In the remainder of this work, we focus exclusively on the evaluation of the on-shell counterparts. Details regarding the computation of the corresponding currents can be found in Ref.~\cite{Bjerrum-Bohr:2018lpz}.

From Eq. \eqref{YMnpointOff} one has,
\begin{align}\label{YM5point} 
&
A_5(1,2,3,4,5)=  
\frac{\sum_M {\cal J}_{3}({\underline{1}},{2},{\underline{\zeta}}^M) {\cal J}_{4}({\underline{\zeta}}'^M,{\underline{3}},4,{5})}{s_{12}}
\nonumber\\
&
+
\sum_{i=3}^4 
\Bigg[ 
\frac{\sum_M 
{\cal J}_{i}({2},{\underline{3}},\Omega_i,{\underline{\kappa}}^M_{i})
{\cal J}_{7-i}({\underline{1}},{\underline{\kappa}}'^M,\Omega'_i,{5})}{s_{2\cdots i}}
\nonumber\\
&
+
2\,
\frac{\sum_L 
{\cal J}_{i}({\underline{2}},{3},\Omega_i,{\underline{\kappa}}^L_{i} )
{\cal J}_{7-i}({\underline{1}},{\underline{\kappa}}'^L_{i},\Omega_i',{5}) 
}{s_{2\cdots i}}\Big|^{1\leftrightarrow 2}\Bigg], 
\end{align}

The non-longitudinal contributions, corresponding to the first two lines, take the same structural form as in the BAS case. Accordingly, employing the independent kinematic invariants specified by the full common set $\mathbf{K}_5$ (see, for instance, Eq.~\eqref{BA5pointOn}), we proceed to evaluate the corresponding on-shell contributions,
\begin{align}
& 1.\,\, \sum_T A_{3}(1,2,\zeta^T) \times A_{4}(\zeta'^T,5,4,3)\Big|_{\mathbf{K}_5},
\nonumber\\
& 2.\,\,  \sum_T  A_{i}(2,3,\Omega_i,\kappa^T_{i}) \times A_{7-i}(1,\kappa'^T_{i},\Omega_i',5) \Big|_{\mathbf{K}_5}, 
\nonumber	
\end{align}
with $i=3,4$, and the transverse completeness relation given in Eq. \eqref{transverse}. 
The explicit computations are carried out in Section \ref{fivepointBCJ}, with particular emphasis on the reconstruction of the five-point BCJ numerators.

In the longitudinal sector, to obtain an on-shell formulation, we must decouple the \mand\ variables associated with the effective ``mass'' of the off-shell legs, \ie $k_{13}^2 = s_{13}$ and $k_{25}^2 = s_{25}$. Since these contributions can be derived from the non-longitudinal terms via the relabeling $1 \leftrightarrow 2$ (see Eq. \eqref{YM5point}), we may employ the same sets of independent kinematic variables used in the preceding on-shell computations, namely,
\begin{equation}
\mathbf{K}_5 \cup \{s_{2\cdots i}\} \text{ for } \,
 \frac{
\sum_L  {\cal J}_{i}({\underline{2}},{3},\Omega_i,{\underline{\kappa}}^L_{i})
 {\cal J}_{7-i}({\underline{1}},{\underline{\kappa}}'^L_{i},\Omega'_i,{5})
 }{s_{2\cdots i}} , \nonumber
\end{equation}
with $i=3,4$.
Following the same procedure used to derive Eq. \eqref{YM3OnLA}, and applying Eqs. \eqref{YMBAnp} and \eqref{YMpBA}, we obtain the on-shell expression for the four-point longitudinal contributions,
\begin{align}\label{YM4OnL}
 &\frac{2}{k_{\kappa_4}^2}
 {\cal J}_{4}({\underline{\kappa}}^l_{4},4,3,{\underline{2}})  
 \nonumber\\
 & 
 =
 \sum_{\delta\in\{ \emptyset , \{4 \} \}} 
 (\epsilon_2\cdot f_\delta \cdot  \epsilon_3)
 {\cal T}_{[\kappa_4,3,\delta ,2]}
 A_{4}( \kappa_{4},4,3,2)\Big|_{\mathbf{K}_5} , \nonumber\\
  &\frac{2}{k_{\kappa'_4}^2}
 {\cal J}_{4}({\underline{\kappa}}'^l_{4},4,5,{\underline{1}})  
 \nonumber\\
 & 
 =
 \sum_{\alpha\in\{ \emptyset , \{4 \} \}} 
 (\epsilon_1\cdot f_\a \cdot  \epsilon_5)
 {\cal T}_{[\kappa'_4,5,\alpha ,1]}
 A_{4}( \kappa'_{4},4,5,1)\Big|_{\mathbf{K}_5} ,
 \end{align}
with $k_{\kappa_4}^2=s_{15}$ and $k_{\kappa'_4}^2=s_{23}$. 

Finally, bringing everything together, as it is detailed in Section \ref{fivepointBCJ} , we arrive at the on-shell formula,
\begin{align}\label{YM5pointon} 
&
A_5(1,2,3,4,5)=
\frac{\sum_T A_{3}(1,2,\zeta^T)\, A_{4}(\zeta'^T,5,4,3)\Big|_{\mathbf{K}_5} }{s_{12}}
\nonumber\\
&
+  
\sum_{i=3}^4\frac{ \sum_T A_{i}(2,3,\Omega_i,\kappa^T_{i}) A_{7-i}(1,\kappa'^T_{i},\Omega'_i,5)\Big|_{\mathbf{K}_5}  }{s_{2\cdots i}}
\nonumber\\
&
-\frac{1}{2}
  \sum_{i=3}^4\sum_{\delta_i, \alpha_i}\! 
 \Big[(\epsilon_2 \cdot f_{\delta_i} \cdot\epsilon_3)(\epsilon_1\cdot f_{\a_i} \cdot \epsilon_5)
 {\cal T}_{[\kappa_{i},3,\delta_i,2]} {\cal T}_{[\kappa'_{i},5,\a_i,1]}
\nonumber\\
&\qquad\,\,
A_{i}(2,3,\Omega_i,\kappa_{i})
 A_{7-i}(1,\kappa'_{i},\Omega'_i,5)\Big|_{\mathbf{K}_5}  
 \Big]\Big|^{1\leftrightarrow 2},
\end{align}
where $\delta_3 = \emptyset$ and $\alpha_3 \in \{\emptyset, \{4\}\}$, while $\delta_4 \in \{\emptyset, \{4\}\}$ and  $\alpha_4 = \emptyset$. Notice that $\delta_i \subseteq \Omega_i$ and $\alpha_i \subseteq \Omega'_i$.

Likewise, the six- and seven-point computations, detailed in \cite{mathematica}, serve to confirm the new on-shell formula proposed below.

\section{New On-Shell Recursive Relation}

The non-longitudinal contributions, corresponding to the first two lines of Eq.~\eqref{YMnpointOff}, are well understood. 
By employing the full common set of independent kinematic invariants $\mathbf{K}_n$, the resulting on-shell expressions are in complete agreement with the contributions obtained from the amputated currents. 
More precisely, one finds,
\begin{align}
&
	\sum_M {\cal J}_{3}({\underline{1}},{2},{\underline{\zeta}}^M) \times
	{\cal J}_{n-1}({\underline{\zeta}}'^M,{\underline{3}},\ldots,,{n}) \nonumber\\
	&\qquad\quad
	=	\sum_T {A}_{3}(1,{2},{\zeta}^T) \times{A}_{n-1}(\zeta'^T,3,\ldots,,{n})\Big|_{\mathbf{K}_n},\nonumber\\
	&\sum_M {\cal J}_{i}\big({2},{\underline{3}},\Omega_i, {\underline{\kappa}}^M_{i} \big)
	\times
{\cal J}_{n+2-i}\big({\underline{1}},{\underline{\kappa}}'^M_{i},\Omega'_i, {n}\big) \nonumber\\
&\qquad\quad
=
\sum_T {A}_{i}\big({2},3,\Omega_i, {\kappa}^T_{i} \big)
\times
{\cal A}_{n+2-i}\big(1,\kappa'^T_{i},\Omega'_i, {n}\big)\Big|_{\mathbf{K}_n}.
\end{align}
These relations have been explicitly verified up to eight-point amplitudes.

The longitudinal contributions, corresponding to the last line in Eq. \eqref{YMnpointOff}, require special consideration. As they can be obtained from the non-longitudinal ones by the relabeling $1 \leftrightarrow 2$, we can use the same sets of independent kinematic variables, \ie $\mathbf{K}_n \cup \{s_{2\cdots i}\}$ for $i = 3, \ldots, n-1$. Now, proceeding as in the previous section, in order to cancel out the spurious poles and obtain an on-shell formulation, we must decouple the \mand\ variables associated with the effective ``mass'' of the off-shell legs. This is achieved by applying the formulae given in Eqs. \eqref{YMBAnp} and \eqref{YMpBA}. Thus, we arrive at the general expressions,
\begin{align}\label{YM4Ons34}
&\frac{2}{k_{\kappa_i}^2} {\cal J}_{i}({\underline{2}},3,\Omega_i,{\underline{\kappa}}^l_{i})= 
\nonumber\\
&   
\sum_{\delta_i \subseteq \Omega_i} (\epsilon_2\cdot F_{\delta_i^t} \cdot \epsilon_3) {\cal T}_{[\kappa_i,3,\delta_i,2]}
A_{i}(2,3,\Omega_i,\kappa_i)\Big|_{\mathbf{K}_n}  ,\nonumber\\
&\frac{2}{k_{\kappa'_i}^2} {\cal J}_{n+2-i}({\underline{1}},{\underline{\kappa}}'^l_{i},\Omega'_i,n)= 
\nonumber\\
&   
\sum_{\alpha_i \subseteq \Omega'_i} (\epsilon_1\cdot F_{\alpha_i^t} \cdot \epsilon_n) {\cal T}_{[\kappa'_i,n,\alpha_i,1]}
A_{n+2-i}(1,\kappa'_{i},\Omega'_i,n)\Big|_{\mathbf{K}_n}  ,
\end{align}
where $k_{\kappa_i}^2=k_{\kappa'_i}^2=s_{2\cdots i}$.

We are now ready to present the on-shell form of the factorization formula in Eq. \eqref{YMnpointOff}, yielding the following recursive relation:
\begin{align}\label{YMnpointOnR}
&
A_n(1,\dots,n)= 
\frac{ \sum_T A_{3}(1,2,\zeta^T) A_{n-1}(\zeta'^T,3,\ldots, n)\Big|_{\mathbf{K}_n}}{s_{12}}  
\nonumber\\
&
+
\sum_{i=3}^{n-1}
\frac{ \mathcal{O}_i \[   
A_{i}(2,3,\Omega_i,\kappa^T_{i})
 A_{n+2-i}(1,\kappa'^T_{i},\Omega'_i, n)\] \Big|_{\mathbf{K}_n} }{s_{2\cdots i}},
\end{align}
with the linear operator $\mathcal{O}_i$ defined by,
\begin{align}\label{Boperator}
&
\mathcal{H}_i\equiv      
\sum_{\delta_i \subseteq \Omega_i \atop 
\a_i \subseteq \Omega'_i
}
(\epsilon_1\cdot F_{\delta_i^t} \cdot \epsilon_3) 
(\epsilon_2\cdot F_{\alpha_i^t} \cdot \epsilon_n)
\mathcal{T}^{(1\leftrightarrow 2)}_{[\kappa_{i},3,\delta_i,2]}
\mathcal{T}^{(1\leftrightarrow 2)}_{[\kappa'_{i},n,\a_i,1]}, 
\nonumber\\
&
\mathcal{O}_i\equiv \sum_T \,\, - \,\, \frac{s_{2\cdots i}}{2} \, \mathcal{H}_i \, , 
\end{align}
\noindent
where 
the superscript $(1 \leftrightarrow 2)$ indicates that the labels 1 and 2 must be exchanged after performing the computations, and 
the transverse completeness relation follows the standard form given in Eq. \eqref{transverse}.

\section{Factorizing BCJ numerators}

Up to this point, we have obtained two novel on-shell recursive factorization relations, Eq. \eqref{BAnpointon} for BAS and Eq. \eqref{YMnpointOnR} for pure YM.  As a byproduct, the on-shell YM formula can be extended to generate BCJ numerators.

To understand how this BCJ factorization arises, we begin with the four-point example. As seen from Eqs. \eqref{YM4pointOn} and \eqref{YMnpointOnR}, it is straightforward to observe that,
\begin{align}\label{YM4BCJOn1} 
&
A_4(1,2,3,4)=  
-\frac{\sum_T N_{(1,2,\zeta^T)}N_{(\zeta'^T,4,3)}}{s_{12}}
\nonumber\\
& 
- 
\frac{1}{s_{14}}
\Big[
\sum_T N_{(1,4,\kappa'^T)}N_{(\kappa^T,2,3)}  
\nonumber\\
&
- \frac{s_{14}}{2}(\epsilon_1\cdot \epsilon_3)(\epsilon_2\cdot\epsilon_4)
{\cal T}^{(1\leftrightarrow 2)}_{[\kappa,3,2]}\,
 {\cal T}^{(1\leftrightarrow 2)}_{[\kappa',4,1]} \,
N_{(1,4,\kappa')} \, N_{(\kappa,2,3)}
\Big] \nonumber\\
&
=-\frac{N_{(1,2,4,3)}}{s_{12}}
-\frac{N_{(1,4,2,3)}}{s_{14}}.
\end{align}
Therefore, 
we can identify the BCJ master numerators,
\begin{align}\label{BCJ4point}
N_{(1,2,4,3)} &= \sum_T N_{(1,2,\zeta^T)}N_{(\zeta'^T,4,3)} \nonumber\\
&
=	
(n_{(124)}\cdot \epsilon_{3}) ,
\nonumber\\
N_{(1,4,2,3)} &= {\cal O} \left[N_{(1,4,\kappa'^T)}N_{(\kappa^T,2,3)}\right]
\nonumber\\
&
=	
(n_{(142)}\cdot \epsilon_{3})- (\epsilon_1\cdot \epsilon_3)(\epsilon_2\cdot\epsilon_4) (k_1\cdot k_4)\, ,
\end{align}
where ${\cal O}$ is the operator,
\begin{equation}
{\cal O}=  \sum_T \,\,- \,\,\, \frac{s_{14}}{2}(\epsilon_1\cdot \epsilon_3)(\epsilon_2\cdot\epsilon_4)
{\cal T}^{(1\leftrightarrow 2)}_{[\kappa,3,2]}\,
 {\cal T}^{(1\leftrightarrow 2)}_{[\kappa',4,1]}.
\end{equation}

This section illustrates, in a simple and explicit manner, how the four-point BCJ numerators emerge from on-shell three-point building blocks.

In the next section, we provide a detailed derivation of the five-point BCJ numerators. A more general formulation will be presented in \cite{inpreparation}.

\section{Five-Point BCJ numerators}\label{fivepointBCJ}

In the previous section, we derived the BCJ master numerators 
$N_{(1,2,4,3)}$ and $N_{(1,4,2,3)}$. 
By relabeling the external legs, the corresponding four-point BCJ numerators 
$N_{(1,2,3,4)}$ and $N_{(1,3,2,4)}$ can be expressed as
\begin{align}\label{BCJ4point}
N_{(1,2,3,4)} &= 
(n_{(123)}\cdot \epsilon_{4}) ,
\nonumber\\
N_{(1,3,2,4)} &= 
(n_{(132)}\cdot \epsilon_{4})
- (\epsilon_1\cdot \epsilon_4)(\epsilon_2\cdot\epsilon_3)\,(k_1\cdot k_3)\, .
\end{align}

Using these master numerators, the four-point amplitude 
$A_4(1,2,3,4)$ can be written as,
\begin{align}\label{YM4BCJOn2} 
A_4(1,2,3,4)
=
\frac{N_{(1,2,3,4)}}{s_{12}}
+
\frac{N_{(1,[2,3],4)}}{s_{14}}\, ,
\end{align}
where we have employed the standard bracket notation,
\begin{equation}
N_{(1,[2,3],4)} \equiv 
N_{(1,2,3,4)} - N_{(1,3,2,4)} .
\end{equation}

In this section, we make use of the two representations of the four-point amplitude presented in Eqs.~\eqref{YM4BCJOn1} and \eqref{YM4BCJOn2}.

Before proceeding with the computation, we note that since all contributions in Eq. \eqref{YMnpointOnR} are on-shell (amplitudes), gauge invariance ensures that the longitudinal components vanish. Therefore, it suffices to employ the transverse completeness relation as, 
\begin{equation}
\sum_T \epsilon_I^{T, \mu} \epsilon_J^{T, \nu} = \eta^{\mu\nu}	,
\end{equation}
where, $I\in\{\zeta,\kappa_i \}$, $J\in\{\zeta',\kappa'_i \}$,  $k_I + k_J = 0$, and $k_I^2 = 0$. 

We now proceed with a detailed evaluation of the contributions obtained in Eq.~\eqref{YM5pointon}. 
Throughout this computation, we use the full common set,
$\mathbf{K}_5 = \{s_{13}, s_{14}, s_{24}, s_{34}\}$, together with the identities,
\begin{align}
-s_{14}-s_{24}-s_{34} &= s_{45} = 2\,k_4\cdot k_5, \nonumber\\
s_{13}+s_{14}+s_{34} &= s_{25} = 2\,k_2\cdot k_5	.
\end{align}
This leads to:
\begin{itemize}
\item ${\bf I}$ contribution:
\end{itemize}
\begin{align}
&
\sum_T A_{3}(1,2,\zeta^T)\, A_{4}(\zeta'^T,5,4,3) \Big|_{\mathbf{K}_5}\nonumber\\
&= 
\frac{ \sum_{T} N_{(1,2,\zeta^T)} {N}_{(\zeta'^T,5,4,3)}}{s_{34}}
+  
\frac{\sum_{T}  N_{(1,2,\zeta^T)} N_{(\zeta'^T,[5,4],3)}}{s_{45}}\, ,
\end{align}
with numerators,
\begin{align}\label{NUM5C1}
\sum_{T} N_{(1,2,\zeta^T)} N_{(\zeta'^T,5,4,3)}
&=
(n_{(1254)}\cdot \epsilon_3), \nonumber\\
\sum_{T} N_{(1,2,\zeta^T)} N_{(\zeta'^T,[5,4],3)}
&=
(n_{(12[54])}\cdot \epsilon_3) \nonumber\\
&+
\frac{s_{14}+s_{24}}{2}
(n_{(12)}\cdot \epsilon_3)(\epsilon_{4}\cdot \epsilon_5).
\end{align}
\begin{itemize}
\item ${\bf II}$ contribution: 
\end{itemize}
\begin{align}
&
\sum_T  A_{4}\big(1,5,4,\kappa'^T_3\big) A_{3}\big(\kappa_3^T,2,3  \big)\Big|_{\mathbf{K}_5}
\nonumber\\
&
 = 
\frac{ \sum_{T} N_{(1,5,4,\kappa'^T_3)} N_{(\kappa_3^T,2,3)}}{ s_{24}+ s_{34}}
+  
\frac{  \sum_{T} N_{(1,[5,4],\kappa'^T_3)} N_{(\kappa_3^T,2,3)} }{ s_{45}}
\, ,
\end{align}
with numerators,
\begin{align}\label{NUM5C2}
\sum_{T} N_{(1,5,4,\kappa'^T_3)} N_{(\kappa_3^T,2,3)}
&
=(n_{(1542)}\cdot \epsilon_3)	\, ,
\nonumber\\
\sum_{T} N_{(1,[5,4],\kappa'^T_3)} N_{(\kappa_3^T,2,3)}
&
= (n_{(1[54]2)}\cdot \epsilon_3) \nonumber\\
&
+ \frac{ s_{14}}{2} (n_{(12)}\cdot \epsilon_3)(\epsilon_{4}\cdot \epsilon_5).
\end{align}
\begin{itemize}
\item ${\bf III}$ contribution: 
\end{itemize}
\begin{align}
&
(-1)\,
\sum_T  A_{3}\big(1,5,\kappa'^T_{4}\big) A_{4}\big(\kappa^T_{4},2,3,4 \big)\Big|_{\mathbf{K}_5}
\nonumber\\
 &= 
\frac{ \sum_{T} N_{(1,5,\kappa'^T_{4})} N_{(\kappa^T_{4},2,4,3)}}{ s_{34}}
+  
\frac{  \sum_{T} N_{(1,5,\kappa'^T_{4})} N_{(\kappa^T_{4},4,2,3)}}{ - s_{24}- s_{34}}
\, ,
\end{align}
with numerators,
\begin{align}\label{NUM5C3}
 \sum_{T} N_{(1,5,\kappa'^T_{4})} N_{(\kappa^T_{4},2,4,3)}
&
=(n_{(1524)} \cdot \epsilon_{3})\, ,
\nonumber\\
 \sum_{T} N_{(1,5,\kappa'^T_{4})} N_{(\kappa^T_{4},4,2,3)}
&
=	(n_{(1542)} \cdot \epsilon_{3}) \nonumber\\
&
+
\frac{s_{24}+ s_{34}}{2}
(n_{(15)} \cdot \epsilon_{3})(\epsilon_{2} \cdot \epsilon_{4}).
\end{align}
\noindent
\begin{itemize}
\item ${\bf IV}$ contribution: 
\end{itemize}
\begin{align}
&
-\frac{1}{2}\,{\cal H}_3 \left[ A_{3}(2,3,\kappa_{3}) A_4(1,5,4,\kappa'_{3}) \right] \Big|_{\mathbf{K}_5}
\nonumber\\
&
=
\frac{1}{2}
  \sum_{\alpha\in\{ \emptyset , \{4 \} \}} 
  (\epsilon_1\cdot\epsilon_3)
 (\epsilon_2\cdot f_\a \cdot \epsilon_5) 
{\cal T}^{(1 \leftrightarrow 2)}_{[\kappa'_{3},5,\a,1]}
\left[
\frac{  N_{(1,5,4,\kappa'_{3})} }{ s_{24}+ s_{34}}\right.  \nonumber\\
&
\left.
\qquad\qquad\qquad
+  
\frac{  N_{(1,[5,4],\kappa'_{3})}  }{ s_{45}}
\right] ,
\end{align}
where,
\begin{align}
{\cal H}_3=
\!\!\!\!\!\!
\sum_{\alpha\in\{ \emptyset , \{4 \} \}} 
(\epsilon_1\cdot\epsilon_3)
(\epsilon_2\cdot f_\a \cdot \epsilon_5) 
{\cal T}^{(1\leftrightarrow 2)}_{[\kappa_{3},3,2]} {\cal T}^{(1\leftrightarrow 2)}_{[\kappa'_{3},5,\a,1]}  \, ,
\end{align}
and with, ${\cal T}^{(1 \leftrightarrow 2)}_{[\kappa_{3},3,2]}A_3(2,3,\kappa_{3}) =-1$.
The numerators are given by,
\begin{align}\label{NUM5C1}
&
\frac{1}{2}
  \sum_{\alpha\in\{ \emptyset , \{4 \} \}} 
  (\epsilon_1\cdot\epsilon_3)
 (\epsilon_2\cdot f_\a \cdot \epsilon_5) 
{\cal T}^{(1 \leftrightarrow 2)}_{[\kappa'_{3},5,\a,1]}  N_{(1,5,4,\kappa'_{3})} \nonumber\\
&
=\frac{1}{2}
(\epsilon_2\cdot \epsilon_5)(\epsilon_1\cdot \epsilon_3) (\epsilon_4\cdot k_{13}) ,
\nonumber\\
&
\frac{1}{2}
  \sum_{\alpha\in\{ \emptyset , \{4 \} \}} 
  (\epsilon_1\cdot\epsilon_3)
 (\epsilon_2\cdot f_\a \cdot \epsilon_5) 
{\cal T}^{(1 \leftrightarrow 2)}_{[\kappa'_{3},5,\a,1]}  N_{(1,[5,4],\kappa'_{3})} \nonumber\\
&
=
-\frac{1}{2}
(\epsilon_1\cdot \epsilon_3)(n_{(54)}\cdot  \epsilon_2) .
\end{align}
\begin{itemize}
\item ${\bf V}$ contribution:  
\end{itemize}
\begin{align}
&
\frac{1}{2}{\cal H}_4\, \left[ A_{4}(\kappa_{4},2,3,4) A_{3}(1,5,\kappa'_{4}) \right] \Big|_{\mathbf{K}_5}
\nonumber\\
&
=
\frac{1}{2}
  \sum_{\delta\in\{ \emptyset , \{4 \} \}} 
  (\epsilon_2\cdot\epsilon_5) 
 (\epsilon_1\cdot f_\delta \cdot \epsilon_3)
{\cal T}^{(1 \leftrightarrow 2)}_{[\kappa_{4},3,\delta,2]}
\left[
\frac{  N_{(\kappa_{4},2,4,3)} }{ s_{34}} \right.
\nonumber\\
&
\qquad\qquad\qquad
\left.
+  
\frac{   N_{(\kappa_{4},4,2,3)}}{ - s_{24}- s_{34}}
\right] ,
\end{align}
where,
\begin{align}
{\cal H}_4=
\!\!\!\!
\sum_{\delta\in\{ \emptyset , \{4 \} \}} 
(\epsilon_2\cdot\epsilon_5) 
 (\epsilon_1\cdot f_\delta \cdot \epsilon_3)
 {\cal T}^{(1 \leftrightarrow 2)}_{[\kappa_{4},3,\delta,2]}
 {\cal T}^{(1 \leftrightarrow 2)}_{[\kappa'_{4},5,1]} 
 \, ,
\end{align}
and with, $ {\cal T}^{(1 \leftrightarrow 2)}_{[\kappa'_{4},5,1]}  A_{3}(1,5,\kappa'_{4}) =-1$.
The numerators are given by,
\begin{align}\label{NUM5C1}
&
\frac{1}{2}
  \sum_{\delta\in\{ \emptyset , \{4 \} \}} 
  (\epsilon_2\cdot\epsilon_5) 
 (\epsilon_1\cdot f_\delta \cdot \epsilon_3)
{\cal T}^{(1 \leftrightarrow 2)}_{[\kappa_{4},3,\delta,2]}
N_{(\kappa_{4},2,4,3)} \nonumber\\
&
=
\frac{1}{2}
(\epsilon_2\cdot \epsilon_5)(n_{(34)}\cdot  \epsilon_1) ,
\nonumber\\
&
\frac{1}{2}
  \sum_{\delta\in\{ \emptyset , \{4 \} \}} 
  (\epsilon_2\cdot\epsilon_5) 
 (\epsilon_1\cdot f_\delta \cdot \epsilon_3)
{\cal T}^{(1 \leftrightarrow 2)}_{[\kappa_{4},3,\delta,2]} 
N_{(\kappa_{4},4,2,3)} \nonumber\\
&
=
\frac{1}{2}
(\epsilon_2\cdot \epsilon_5)(\epsilon_1\cdot \epsilon_3) (\epsilon_4\cdot k_{13}).
\end{align}

Finally, by assembling all these on-shell contributions in Eq. \eqref{YM5pointon}, we arrive at the expected result,
\begin{widetext} 
\begin{align}\label{YM5pointOnR}
A_5(1,2,3,4,5)&= 
\frac{(n_{(1254)}\cdot \epsilon_3)+ (\epsilon_2\cdot \epsilon_5)(n_{(34)}\cdot  \epsilon_1) (k_{1}\cdot k_{2}) }{s_{12}\, s_{34}}+ \frac{(n_{(1[54]2)}\cdot \epsilon_3) + (n_{(12)}\cdot \epsilon_3)(\epsilon_{4}\cdot \epsilon_5) (k_1\cdot k_4) }{s_{23}\, s_{45}}
\nonumber\\
&
+
\frac{
(n_{(12[54])}\cdot \epsilon_3)+(\epsilon_{4}\cdot \epsilon_5)(n_{(12)}\cdot \epsilon_3)(k_{12}\cdot k_4 )
- (\epsilon_1\cdot \epsilon_3)(n_{(54)}\cdot  \epsilon_2) (k_1\cdot k_2) 
}{s_{12}\, s_{45}} +
\frac{(n_{(1524)} \cdot \epsilon_{3})}{s_{234}\, s_{34}}
\nonumber\\
&
+
\frac{1}{s_{23}\, s_{234}}
\left\{(n_{(1542)}\cdot \epsilon_3) 
\left[\frac{1}{\left(1-\frac{s_{23}}{s_{15}}\right)}+\frac{1}{\left(1-\frac{s_{15}}{s_{23}}\right)}\right] 
- 
(n_{(15)} \cdot \epsilon_{3})(\epsilon_{2} \cdot \epsilon_{4})(k_2\cdot k_3)	
\right\}
\nonumber\\
&=
\frac{N_{(1,2,5,4,3)}}{s_{12}\, s_{34}} 
+  
\frac{N_{(1,2,[5,4],3)}}{s_{12}\,s_{45}}
+
\frac{N_{(1,[5,4],2,3)}}{s_{23}\,s_{45}}
+  
\frac{N_{(1,5,4,2,3)}}{s_{23}\,s_{234}}
+  
\frac{N_{(1,5,2,4,3)}}{s_{234}\,s_{34}}
\, ,
\end{align}
\end{widetext}
from which the BCJ numerators can be directly read off:
\begin{align}
N_{(1,2,5,4,3)} & = 
(n_{(1254)}\cdot \epsilon_3)+ (\epsilon_2\cdot \epsilon_5)(n_{(34)}\cdot  \epsilon_1) (k_{1}\cdot k_{2}),
\nonumber\\
N_{(1,2,[5,4],3)} & = 
(n_{(12[54])}\cdot \epsilon_3)+(\epsilon_{4}\cdot \epsilon_5)(n_{(12)}\cdot \epsilon_3)(k_{12}\cdot k_4 ) \nonumber\\
&
- (\epsilon_1\cdot \epsilon_3)(n_{(54)}\cdot  \epsilon_2) (k_1\cdot k_2), 
\nonumber\\
N_{(1,[5,4],2,3)} & = 
(n_{(1[54]2)}\cdot \epsilon_3) + (n_{(12)}\cdot \epsilon_3)(\epsilon_{4}\cdot \epsilon_5) (k_1\cdot k_4),
\nonumber\\
N_{(1,5,4,2,3)} & = 
(n_{(1542)}\cdot \epsilon_3)-(n_{(15)} \cdot \epsilon_{3})(\epsilon_{2} \cdot \epsilon_{4})(k_2\cdot k_3)	,
\nonumber\\
N_{(1,5,2,4,3)}	& = (n_{(1524)} \cdot \epsilon_{3}). \nonumber\\
\end{align}

The above BCJ numerators are in complete agreement with those obtained from the construction of Ref.~\cite{Du:2017kpo}, evaluated with reference ordering ${\rm RO} = \{1,5,2,4,3\}$.

\section{Conclusions} 

We introduced a new method for deriving on-shell recursive relations for scattering amplitudes, using three-point amplitudes as fundamental building blocks. 
 
By selecting an appropriate set of independent kinematic variables ({\it common set}), we showed that each off-shell factorization contribution is independent of the \mand\ variable associated with the effective ``mass'' of the off-shell leg, \eg $\partial_{s_{23\cdots i}}(\mathcal{J}_{i} \times \mathcal{J}_{n+2-i}) = 0$. Exploiting this translational symmetry in kinematic space, we gauge-fix this \mand\ variable to zero, \ie $s_{2\cdots i} = 0$, thereby projecting each contribution onto its on-shell counterpart ({\it amplitude}). While this work focused on ${\rm Tr}(\phi^3)$ and pure Yang-Mills theories, we are currently extending the method to a wider class of theories, including those with fermionic degrees of freedom, non-linear sigma model, higher-derivative theories (such as $(DF)^2$ and $F^3$),
 and exploring the role of amputated currents in universal splitting processes \cite{inpreparation,Cao:2024gln,Cachazo:2021wsz,Arkani-Hamed:2024fyd,Zhang:2024iun,Zhou:2024ddy,Feng:2025ofq,Azevedo:2025vxo,Zhang:2026dcm}.

Several key properties, such as gauge invariance, soft behavior, factorization, and cyclic symmetry, can be readily confirmed from Eqs.~\eqref{BAnpointon} and \eqref{YMnpointOnR}. These checks  providing an important consistency check for the correctness of our final results.

Through a detailed analytic study of the ${\rm Tr}(\phi^3)$ theory, we showed that the spurious poles appearing in the recursive expression of Eq. \eqref{BAnpointon} cancel out, following a mechanism analogous to that of BCFW recursion with shifted momenta $k_1(z) = k_1 + zq$ and $k_2(z) = k_2 - zq$, where $q \cdot k_1 = q \cdot k_2 = q^2 = 0$. Moreover, our approach provides a simple and systematic way to understand and compute the boundary contributions associated with this type of deformation.

Remarkably, the same cancellation mechanism also appears in pure Yang-Mills amplitudes, as shown in Eq.~\eqref{YM5pointOnR}. However, establishing relations analogous to those in Eqs.~\eqref{fidentity}, \eqref{sidentity}, and \eqref{Id5point} is considerably more involved, since in pure Yang-Mills the polarization vectors must also be consistently deformed. A detailed investigation of these aspects will be presented in future work \cite{inpreparation}.

One of the central ingredients of our construction is the ability to rewrite longitudinal contributions in a purely on-shell form. As shown by the authors in \cite{Bjerrum-Bohr:2018lpz}, contact terms in pure Yang-Mills can be expressed as longitudinal contributions arising from the product of two smaller amputated currents (see Eq.~\eqref{YMnpointOff}). This observation suggests that, in order to obtain an on-shell prescription, one must factorize the effective ``mass'' associated with the longitudinal leg from these contributions. Schematically, this takes the form,
\begin{equation}\label{JandA}
{\cal J}_m\Big|_{\epsilon_\kappa \to k_\kappa} = \,\,k_\kappa^2 \,\,\,\sum\, A_m \,  .
\end{equation}

Since the polarization vector of the longitudinal leg is removed from the amputated current ${\cal J}_m$, the right hand side of Eq.~\eqref{JandA} must involve scalar degrees of freedom. This naturally leads to
\begin{equation}
{\cal J}_m\Big|_{\epsilon_\kappa \to k_\kappa} = \,\,k_\kappa^2 \,\,\,\sum \, A^{\rm YM \oplus \phi^3}_m  \, ,
\end{equation}
where the amplitudes now include both gluons and scalar particles.

Finally, in order to express the result entirely in terms of pure gluon amplitudes, it is convenient to introduce the well-known transmutation operators \cite{Cheung:2017ems}, which map gluonic states into scalar ones. Schematically, this relation can be written as
\begin{equation}
A^{\rm YM \oplus \phi^3}_m = {\cal T} \cdot A^{\rm YM}_m , 
\end{equation}
where ${\cal T}$ is a transmutation operator.
This constitutes a highly nontrivial analysis, in which we make use of the identity given in Eq.~\eqref{YMBAnp}, as proposed in \cite{Dong:2021qai,Zhou:2022orv}.

As a natural consequence, this new recursive structure facilitates the reconstruction of BCJ numerators in a simple factorized form  (several examples will be presented in Ref. \cite{inpreparation}). This framework not only provides new insights into the structure of gauge theory amplitudes but may also open avenues for future developments in gravitational theories via the double-copy construction, or in Cosmological scattering equations \cite{Eberhardt:2020ewh,Roehrig:2020kck,Diwakar:2021juk,Gomez:2021qfd}.

Finally, while our analysis was carried out within the DC formalism using the specific gauge fixing $(pqr|m) = (n12|3)$, the construction can be straightforwardly extended to an arbitrary gauge choice. 
In such a case, 
the matrix given by the Mandelstam invariants, $s_{ab}$, can be arranged according to the structure,
\begin{align}
s_{ab} \,=\,
\Bigg(
\begin{matrix}
q & r & m & \Omega_i \cup \Omega'_i & p
\end{matrix}
\Bigg),
\end{align}
where 
$\Omega_i \cup \Omega'_i=\{1,2,\ldots,n\}\setminus\{p,q,r,m\}$. Therefore, the full common set $\mathbf{K}_n$ keeps the same polygonal form as in Eq.~\eqref{Matrixnp}.

\begin{acknowledgements}  

Numerous discussions with N. Emil J. Bjerrum-Bohr are gratefully acknowledged. We also thank to C. Lopez-Arcos, R. Lipinski Jusinskas and J. Raeymaekers for useful comments.
This work is supported by the GA\v{C}R grant 25-16244S from the Czech Science Foundation.
\end{acknowledgements}

\end{document}